\documentclass[twocolumn,tighten, times]{aastex63}
\usepackage{amsmath}
\usepackage{lipsum}
\usepackage{csquotes}

\received{February 24, 2021}
\shorttitle{Spectral modeling of PKS 0903-57}
\shortauthors{Mondal et al.}

\begin{document}

\vspace*{-1.5cm}
\title{Spectral Modeling of Flares in Long Term Gamma-Ray Light Curve of PKS 0903-57}

\email{skmondal@rri.res.in}

\author[0000-0003-2445-9935]{Sandeep Kumar Mondal}
\affiliation{Raman Research Institute, C. V. Raman Avenue, Sadashivnagar, Bangalore: 560080, India}


\author[0000-0002-1173-7310]{Raj Prince}
\affiliation{Center for Theoretical Physics, Polish Academy of Sciences, Al.Lotnikow 32/46, 02-668, Warsaw, Poland}

\author[0000-0002-1188-7503]{Nayantara Gupta}
\affiliation{Raman Research Institute, C. V. Raman Avenue, Sadashivnagar, Bangalore: 560080, India}

\author[0000-0002-9526-0870]{Avik Kumar Das}
\affiliation{Raman Research Institute, C. V. Raman Avenue, Sadashivnagar, Bangalore: 560080, India}



\begin{abstract}
{A detailed study of the BL Lacertae PKS 0903-57 has been done for the first time with 12 years of Fermi Large Area Telescope data. We have identified two bright gamma-ray flares in 2018 and 2020. Many sub-structures were observed during multiple time binning of these flares. We have performed detailed temporal and spectral study on all the sub-structures separately. A single-zone emission model is used for time-dependent leptonic modeling of the multi-wavelength spectral energy distributions. 
Our estimated values of variability time scale, magnetic field in the emission region, jet power obtained from leptonic modeling of PKS 0903-57 are presented in this work. Currently, we have a minimal number of observations in X-rays and other bands. Hence, more simultaneous multi-wavelength monitoring of this source is required to have a better understanding of the physical processes happening in the jet of the blazar PKS 0903-57.}

\end{abstract}



\section{Introduction} \label{sec:intro}

Active Galactic Nucleus (AGN) is the central core of an active galaxy, which is much brighter than its host galaxy and highly variable. It is thought that at the centre of an active galaxy there is a super-massive black hole, which accretes matter and forms an accretion disk around the core. AGN emits jet along its polar direction i.e. perpendicular to its accretion disk, which are collimated, narrow beams of highly relativistic particles. The nonthermal emission from the highly relativistic particles  is detected in radio to gamma-ray frequency.
\par
 Depending on the orientations of the jets of AGNs with respect to the observer's line of sight, AGNs have been classified into several classes. If the jet is oriented close to the line of sight of the observer \citep{Urry_1995} then it is called a blazar. Moreover, blazars have two sub-classes: flat-spectrum radio quasars (FSRQs) and BL Lacertae (BL Lac). FSRQs show characteristic spectral lines whereas BL Lacs show featureless spectra or very weak spectral lines. Blazars are highly variable; their variability time varies from minutes \citep{2007ApJ...664L..71A} to year \citep{2013MNRAS.436.1530R} scale across the whole electromagnetic spectrum. They are also found to be the sources of high-energy gamma-rays in the universe. The
  mechanisms of particle acceleration in jet, the magnetic field structure in jet, the underlying causes of variability in jet emission over short and long time scales are not yet well understood. Multi-wavelength data analysis and modeling of jet emission are necessary to understand the physics of these objects.

PKS 0903-57 is a BL Lac \citep{refId0} type object also known as 3FGL J0904.8-5734 or 4FGL J0904.9-5734, located at redshift z= 0.695 \citep{Thompson_1990} with RA= 136.222 deg or  09h04m53.1790s \& DEC=-57.5849 deg or -57d35m05.783s \citep{Fey_2004}. 

 PKS 0903-57 was studied for the first time in 1987 \citep{10.1093/mnras/227.3.705} and classified as a quasar by FST (Fleurs Synthesis Telescope). In 1990 based on the optical brightness, 37 PKS sources ( detected by the Parkes Observatory, Parkes 2700 MHz survey) were observed and PKS 0903-57 was one of them; mentioned as `0903-573'. They classified it as `Seyfert I'. As they were not sure about the classification, it was listed among the `misidentified sources'. Later, in both of the Fermi 3FGL \citep{latcollaboration2015fermi} \& 4FGL \citep{Abdollahi_2020} catalogs, this source has been classified as `BCU' (`Blazar candidates of uncertain type'). In `Simbad' it is classified as `BL Lac type object' as suggested by \cite{refId0} . In the period between August 2008 to the beginning of 2020, two enhanced $\gamma$-ray activity states were reported, though they were much dimmer compared to the flares detected during March-April of 2020. Along with Fermi Large Area Telescope (\textit{Fermi}-LAT) other observatories like ATCA (Australia Telescope Compact Array), Swift, AGILE (Astrorivelatore Gamma ad Immagini LEggero), DAMPE (Dark Matter Particle Explorer), HESS (High Energy Stereoscopic System) also observed this source during March-April 2020. Following are the details of the alerts, from where we get a brief history of its activity.

On 22nd June 2015, \textit{Fermi}-LAT detected $\gamma$-ray flare from this source \citep{2015ATel.7704....1C} with daily average flux  (1.2$\pm$0.3)$\times$10$^{-6}$ ph cm$^{-2}$ s$^{-1}$ above 100 MeV, which was 30 times higher than its average flux in 3FGL catalog. 

Again, \textit{Fermi}-LAT detected GeV $\gamma$-ray flare from this source on 14th May 2018 \citep{2018ATel11644....1C} with daily average flux (2.2$\pm$0.2)$\times$10$^{-6}$ ph cm$^{-2}$ s$^{-1}$ above 100 MeV, about 55 times higher than its flux reported in the 3FGL catalog.

In 2020, very high $\gamma$-ray flux was detected by \textit{Fermi}-LAT; this is the highest $\gamma$-ray flux ever detected from this source. On 28th March 2020, an elevated $\gamma$-ray flux with two GeV photons (E$>$10 GeV) with daily average flux  (3.8$\pm$0.4)$\times$10$^{-6}$ ph cm$^{-2}$ s$^{-1}$ above 100 MeV was observed  \citep{2020ATel13599....1M}. This time the $\gamma$-ray flux was about 60 times higher than the average flux reported in the 4FGL catalog. This was the third time \textit{Fermi}-LAT detected such enhanced $\gamma$-ray activity from this source. This flare was also reported in the LAT GCN 1585493148.

 AGILE reported enhanced $\gamma$-ray activity from the same source on 1st April 2020 \citep{2020ATel13602....1L}. 
 
 \textit{Fermi}-LAT also reported very-high energy $\gamma$-ray emission from PKS 0903-57 on 1st April 2020 \citep{2020ATel13604....1B}. Preliminary analysis of \textit{Fermi}-LAT data reported several high-energy ($>$10 GeV) photons which were positionally consistent with this source. It was found that the association of those high-energy photons with this source was highly probable. Amongst them a 106 GeV photon was detected on 31st March 2020 at 13:56:27.000 UTC. This was the first evidence of VHE (Very-High Energy) $\gamma$-ray emission from PKS 0903-57 by \textit{Fermi}-LAT. 
 
 On 13th April 2020, first time HESS reported the detection of very-high energy $\gamma$-ray \ during a follow-up observation from the intermediate BL Lacertae object PKS 0903-57 \citep{2020ATel13632....1W}. 
 
 ATCA monitored this source periodically. On 15th April 2020, ATCA released a report on the recent activity of this source \citep{2020ATel13638....1S}. They observed this source on 2nd April 2020 in six radio bands with a duration of 10 mins in each band and reported the fluxes in each band. 

DAMPE reported about the detection of GeV $\gamma$-rays from the source PKS 0903-57 on 17th April 2020 \citep{2020ATel13643....1D} with daily average flux  $\sim$(5.9$\pm$2.3)$\times$10$^{-7}$ ph cm$^{-2}$ s$^{-1}$.

The underlying mechanism of flux variability of the blazars is still unknown to the community. Many models have been proposed to explain the variability in short time scale but the models are highly flare-dependent and in some cases also source dependent. The total spectral energy distributions (SEDs) of blazars show two hump-like structures. The first hump covering the optical-UV and soft X-ray part of the EM spectrum is produced by synchrotron emission of relativistic leptons in the jet's emission region. The second hump covers a broader energy range, from soft X-ray to very high energy $\gamma$-ray. 
In the leptonic model, the second hump is explained by Inverse Compton (IC) emission of relativistic leptons.
The seed photons for the IC emission may be the synchrotron photons emitted by the relativistic leptons; in this case, the IC emission is known as Synchrotron Self-Compton (SSC) emission. This mechanism explains the second humps in the SEDs of most of the BL Lacs. In FSRQs, the IC emission may happen with the seed photons outside the jet, which is known as External Compton (EC) emission. Since our source is a BL Lac, we focus on the SSC emission to explain the second hump of the SEDs.
\par
 In our 12 years long \textit{Fermi}-LAT $\gamma$-ray data analysis we have identified two flares: Flare-1 \& Flare-2. Further smaller binning of the $\gamma$-ray light curve reveals the sub-structures prominently. Flare-1 has one sub-structure, consisting of two phases. Flare-2 has two sub-structures: Flare-I and Flare-II; the first one has five phases and the later one has three phases. These phases consist of preflare, flares and postflare states. The flaring phases have been fitted with a sum of exponential equations to calculate the rising \& decay time of the peaks of the phases. Thereafter we have calculated the $\gamma$-ray variability time which is found to be hour scale. We have fitted the $\gamma$-ray SEDs of different phases with different models, PowerLaw (PL), LogParabola (LP), BrokenPowerLaw (BPL) and PowerLaw Exponential Cutoff (PLEC) to find out the model which represents the data best. We do not find any specific hardening or softening pattern in the fitted spectrum. 
 On the basis of the maximum likelihood analysis, LogParabola is the best-fitted model which we have used in modeling the multi-wavelength SED with the help of a time-dependent code. Our results show that one zone leptonic model is sufficient to model the multi-wavelength SED. For better understanding of the physical processes more simultaneous multi-wavelength data is required. 
 
\par
In this paper, we have studied the $\gamma$-ray data from 4th August 2008 to 6th January 2021. After identifying significant flares in the $\gamma$-ray data, we have included the multi-wavelength data from several instruments and modeled the flaring phases. In \autoref{Sec:Sec2_Data_Analysis}, we have discussed the multi-wavelength data analysis; in \autoref{Sec:Sec3_Flaring_States} we have identified the flares and their sub-structures from the  $\gamma$-ray light curve.
 In \autoref{Sec:Sec4__Phase_Detection_Fitting}, we have discussed the method of identification of different phases of the flares \& the fitting of the $\gamma$-ray light curve with a functional form to compute the rising and decay timescale of the flaring phases.
 In \autoref{Sec:Sec5_Flare_Descriptions}, we have discussed about the $\gamma$-ray flares, their sub-structures \& phases in detail. 
 In \autoref{Sec:Sec6_GammaR_SED}, we have discussed the fitting of the $\gamma$-ray SEDs with different functional forms e.g. PowerLaw (PL), LogParabola (LP), BrokenPowerLaw (BPL) and PowerLaw with Exponential Cutoff (PLEC). 
 In \autoref{Sec:Sec7_MW_Study}, we have done time-dependent modeling of the multi-wavelength SEDs with `GAMERA' and calculated the total jet power required in our model. We have discussed our results in \autoref{Sec:Sec8_Discussion} and the conclusion is given in \autoref{Sec:Sec9_Conclusion}.
 \par

\section{Data Analysis}
\label{Sec:Sec2_Data_Analysis}
	\subsection{\textit{Fermi}-LAT Data Analysis}
	\textit{Fermi}-LAT is one of the two instruments of Fermi Gamma-Ray Space Telescope \citep{Atwood_2009}. It is an imaging, pair-conversion, high-energy gamma-ray telescope that can detect photons of energy between 20 MeV to more than 1 TeV, whose field of view is 2.7 sr at 1 GeV and above \citep{Abdollahi_2020}. It scans the whole sky every three hours. It was launched in June 2008 in the near-earth orbit and still in operation.  PKS 0903-57 was continuously monitored by \textit{Fermi}-LAT from 4th August 2008, 15:43:36 UTC and also listed in their regularly monitored source-list. The Pass 8 \textit{Fermi}-LAT $\gamma$-ray data of PKS 0903-57 was extracted from Fermi Science Support Center (FSSC) data server\footnote{\url{https://fermi.gsfc.nasa.gov/ssc/data/access/lat/msl_lc/source/PKS_0903-57}} for a period of more than 12 years (August 2008 to January 2021) and the analysis was done with  Fermi Science Tools software package version v11r5p3 \citep{Fermi_Sc_Tools}, following the `Unbinned Likelihood analysis' method. We have used Pass 8 data where the photon-like events are classified as evclass=128 (the \textit{Fermi}-LAT collaboration recommended to use the `SOURCE' event class for relatively small regions of interest ($<$25 deg) \citep{bruel2018fermilat}. We have used `P8R3\_SOURCE' event class for which `evclass' has to be set to a value 128)\footnote{\url{https://fermi.gsfc.nasa.gov/ssc/data/analysis/documentation/Cicerone/Cicerone_Data/LAT_DP.html}} and evtype=3 (each event class includes different event types which allows us to select events based on different criteria. The standard value of `evtype' is 3 which includes all types of events i.e. front and back section of the tracker (denoted by FRONT+BACK), for a given class.). We have extracted the \textit{Fermi}-LAT Gamma-Ray data from FSSC considering a search radius of 20$^{\circ}$ around the source PKS 0903-57. During the data preparation we have selected `Region of Interest (ROI)' of 10$^{\circ}$, as suggested in Fermi's Data Preparation page \footnote{\url{https://fermi.gsfc.nasa.gov/ssc/data/analysis/documentation/Cicerone/Cicerone_Data_Exploration/Data_preparation.html} }. As per the analysis method, NewMinuit should be converged. We have followed the steps mentioned in Fermi data analysis manual and finally NewMinuit has converged for ROI=7$^{\circ}$. Further study is done for considering the photons from a ROI of 7$^{\circ}$ around the source, and the maximum zenith angle of 90$^{\circ}$ was chosen to avoid earth limb contamination in our analysis.
\par
Moreover, we have used  \textit{Fermi}-LAT fourth source catalog (4FGL) \citep{Abdollahi_2020} and the galactic diffuse emission model (gll$\_$iem$\_$v07.fits) and extra-galactic isotropic diffuse emission model (iso$\_$P8R3$\_$SOURCE$\_$V2$\_$v1.txt) to build the model xml file. After selection of the events based on the cuts; good time interval data, livetime, exposure map, diffuse response of the instrument have been computed eventually for each event with the instrument response function (IRF) `P8R3$\_$SOURCE$\_$V2$\_$v1'. The model xml file would have many sources within the ROI and the likelihood analysis optimizes the spectral parameters of all the sources. The model xml file also has sources outside the ROI which are generally fixed to their 4FGL catalog values.
For localization of the source a quantity, `Test Statistics' (TS) is generally computed, defined as,
	\begin{equation}
TS =-2log \frac{L_{0}}{L_{1}}
\end{equation}
where, $L_{0}$ and $L_{1}$ are the maximum likelihood value for a model without (null hypothesis) and with a point like source at the position of the source respectively. Larger is the TS value; higher is the probability of the presence of the source. To generate the light curve, we have fixed all the parameters of all the sources in our radius of interest (ROI) except our source of interest from the fourth \textit{Fermi}-LAT catalog (4FGL). In this paper, we have generated $\gamma$-ray light curve in five different time bins: 7-day, 1-day, 12-hour, 6-hour and 3-hour and subsequently generated SEDs of different activity periods.

\subsection{Swift XRT and UVOT Data Analysis}
   Swift is a multi-wavelength space-based observatory with three instruments onboard: BAT, XRT, UVOT \citep{Burrows_2005}. It observes the sky in hard X-ray, soft X-ray, Ultraviolet and Optical wavebands. PKS 0903-57 was monitored by Swift during its flaring states. The details of the observations are tabulated in \autoref{tab:SWIFT Observation IDs}. Nearly, 15 observations are found corresponding to the detected $\gamma$-ray flares.
   
    In Swift-XRT data, we have used clean event files corresponding to Photon-Count mode (PC mode), which we obtained using a task `xrtpipeline' version 0.13.5. Calibration file (CALDB), version 20190910 and other standard screening criteria have been applied to the cleaned data. A radius of interest (ROI) of 20-30 pixel has been considered to mark the source region, the radius of the background region is also the same, but it is far away from the source region. With the help of `xselect' tool, we have selected source region \& background region and saved the spectrum files of the corresponding region. Then `xrtmkarf' and `grppha' tools have been used to generate ancillary response file and group the spectrum file with the corresponding response matrix file; thereafter `addspec' and `mathpha' have been used. The SEDs corresponding to different flaring phases have been obtained. Thereafter, the spectra have been modeled with xspec \citep{1996ASPC..101...17A} (Version 12.11.0) tools. During fitting, we have considered neutral hydrogen column density, n\textsubscript{H}=2.6$\times10^{21}$ cm$^{-2}$ \footnote{\url{https://heasarc.gsfc.nasa.gov/cgi-bin/Tools/w3nh/w3nh.pl}}. These X-ray SEDs have been shown in the multi-wavelength SEDs corresponding to their flaring phases.
    
    PKS 0903-57 was also monitored by Swift-UVOT (Ultraviolet/Optical Telescope) in all six filters: U (3465 \AA), V (5468 \AA), B (4392 \AA), UVW1 (2600 \AA), UVM2 (2246 \AA) and UVW2 (1928 \AA). The source region has been extracted from a region of 5 arcsec around the source, keeping the source at the centre of the circle. The background region has been taken $\sim$3 times larger than the source region far away from the source region. Using `uvotsource' tool, we have extracted the source magnitude. This magnitude doesn't consider the galactic absorption, so it has been corrected. As there is no documentation from where we can collect the extinction value for this source, we have used the extinction value using a python module `extinction' \footnote{\url{https://extinction.readthedocs.io/en/latest/}} corresponding to all the Swift-UVOT filters for this source. We have considered Fitzpatrick (1999) \citep{Fitzpatrick_1999} dust extinction function for R\textsubscript{V}=3.1. Following are the values of the extinction coefficients of different Swift-UVOT wavebands that have been used here; V: 0.986, B: 1.311, U: 1.591, UVW1: 2.126, UVM2: 2.958, UVW2: 2.614.
    Subsequently, the corrected magnitudes have been converted into flux by using zero point correction \citep{Breeveld_2011}.

\section{Flaring States of PKS 0903-57}
\label{Sec:Sec3_Flaring_States}
\begin{figure}[h]
\centering
\includegraphics[width=0.94\textwidth]{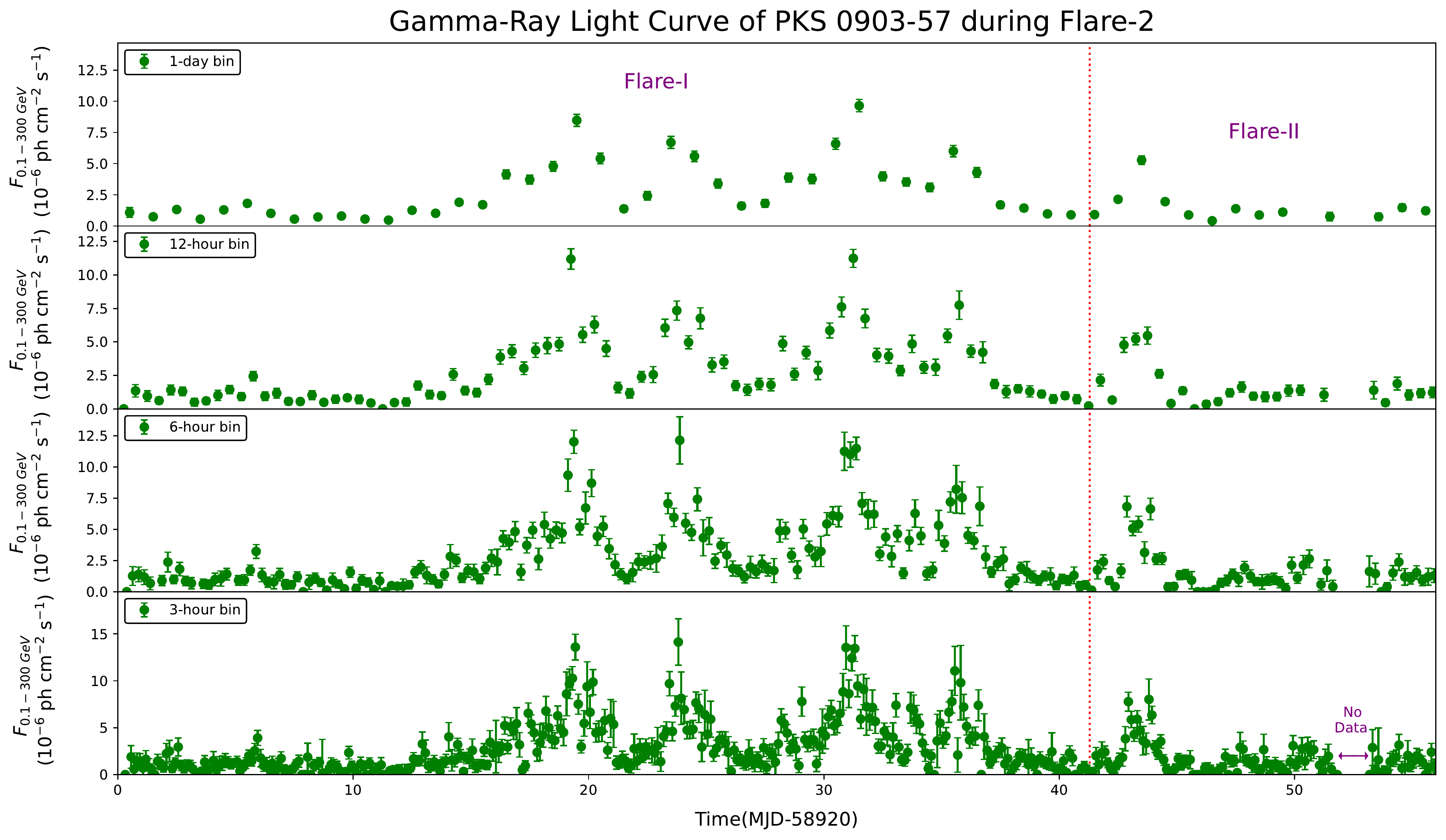}
\caption{$\gamma$-ray light curve of PKS 0903-57 during Flare-2 (MJD 58920.0-58976.0). Two sub-structures of Flare-2: Flare-I \& Flare-II.}
\label{Fig:Flare-I_&_II_Combo}
\end{figure}

      We have analysed the $\gamma$-ray light curve of PKS 0903-57 observed over 12 years in different time bins. \autoref{Fig:7D_Whole} shows 7-day binning of $\gamma$-ray light curve of this source, which was observed by \textit{Fermi}-LAT from MJD 54682.65 (4th August 2008; 15:43:36 UTC) to 59220 (6th January 2021; 00:00:00 UTC). From \autoref{Fig:7D_Whole}, we have identified two flaring states (denoted by a pair of vertical red-dotted lines for each flaring states). We have indicated these two flaring states as Flare-1 and Flare-2, which were observed between MJD 58216.5 to 58230 and MJD 58920 to 58976 respectively. Our work is focused on the flaring states; hence a detailed analysis has been carried out on the flaring states only. 
      Within a larger flare, there are smaller flares with preflare and postflare states before and after them. A sub-structure consists of multiple phases of flare, preflare and postflare.
      We have studied the flares in 1-day time bin to detect their sub-structures and different phases, thereafter we have analysed them in 12-hour, 6-hour and 3-hour time bin to detect the sub-structures and phases more precisely. Flare-1 has only one sub-structure (\autoref{Fig:Flare-1_Multi_Bin}) whereas Flare-2 has two sub-structures (\autoref{Fig:Flare-I_&_II_Combo}), labeled as: Flare-I \& Flare-II.
      
        For some phases, the error bars on the data points in their 3-hour binned $\gamma$-ray light curves are larger compared to those in their 6-hour binned $\gamma$-ray light curves. However, in their 6-hour binned $\gamma$-ray light curves, the flaring segments have comparatively fewer data points compared to those in their 3-hour binned $\gamma$-ray light curves. 
       If the number of data points is not sufficient it is hard to determine the position of the peak in the light curve during a flare. Hence, we have used the 3-hour binned $\gamma$-ray light curves in our study, although their data points have comparatively larger error bars in some cases.

      Throughout the paper, the $\gamma$-ray fluxes have been reported in 10$^{-6}$ ph cm$^{-2}$ s$^{-1}$ unit (in the text; in case of figures we have mentioned the unit in the bracket).

\section{Method of Identification of Different Activity Phases \& Temporal Evolution of Gamma-Ray Light Curve during the Flares }
\label{Sec:Sec4__Phase_Detection_Fitting}
\begin{figure}[h]
\centering
\includegraphics[width=0.91\textwidth]{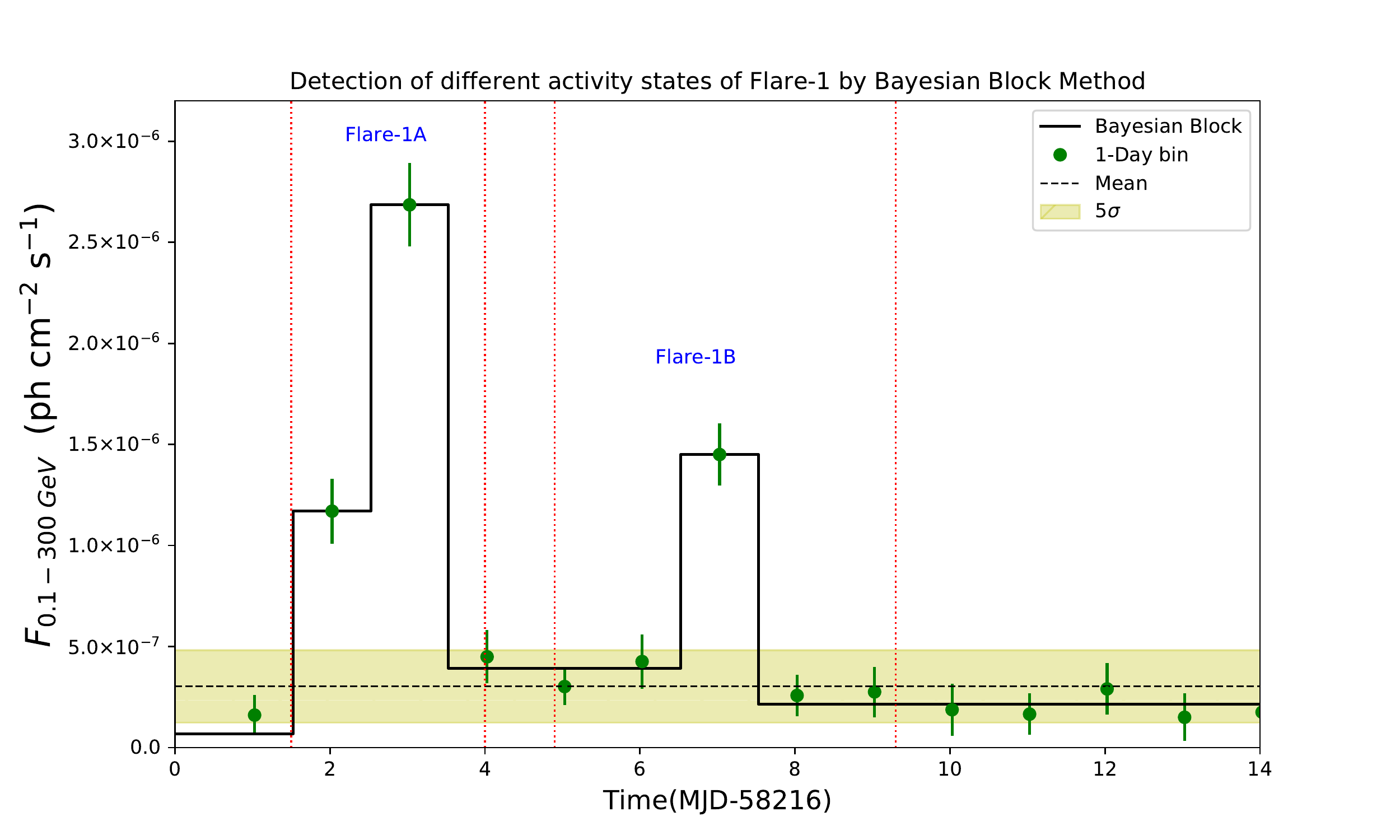}
\caption{`Flare' detection with the help of Bayesian Block method during `Flare-1'.}
\label{Fig:Flare-1_Bayesian_Block}
\end{figure}

\begin{figure}[h]
\centering
\includegraphics[width=0.94\textwidth]{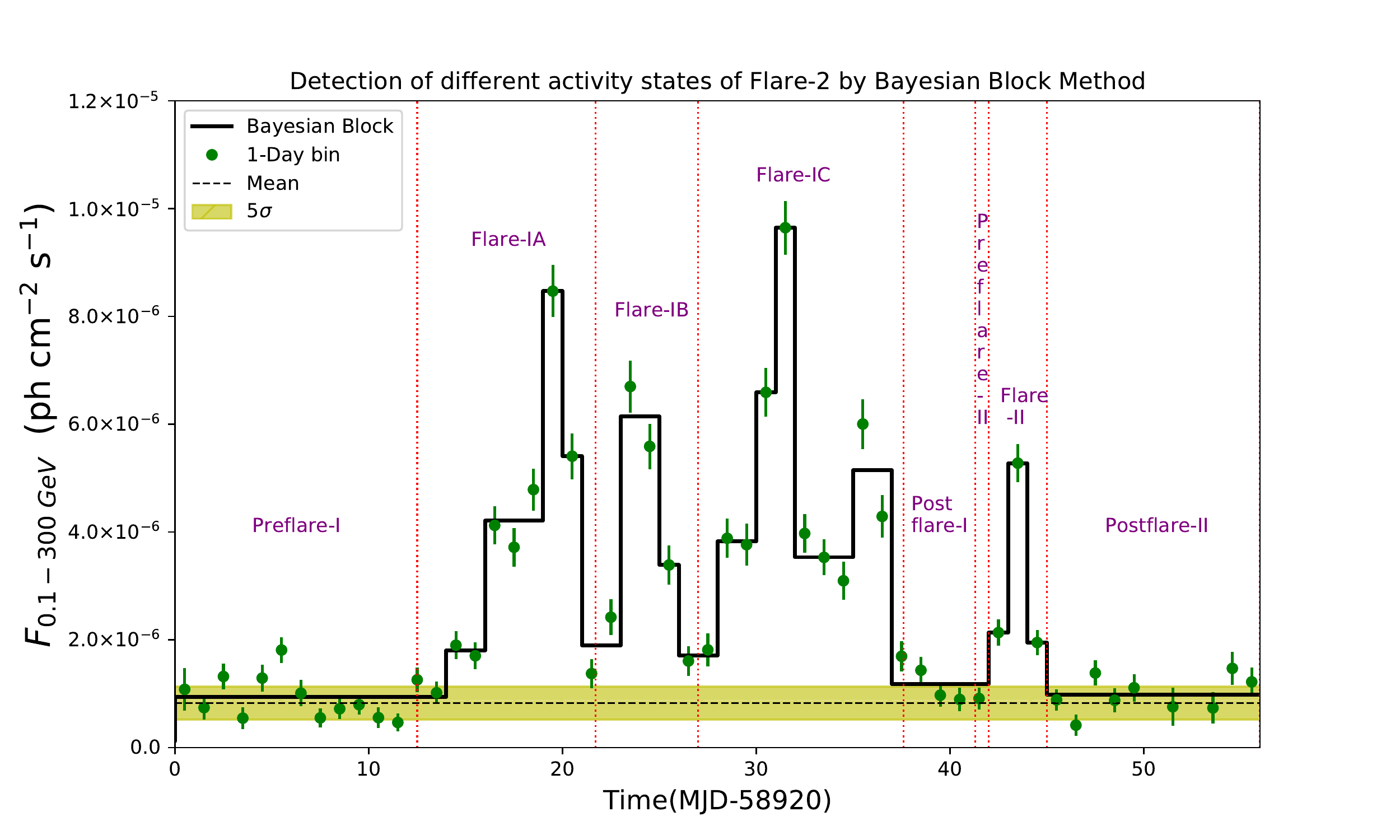}
\caption{`Flare' detection with the help of Bayesian Block method during `Flare-2'.}
\label{Fig:Flare-2_Bayesian_Block}
\end{figure}

\begin{figure}[h]
\centering
    \begin{minipage}{0.49\textwidth}
     \includegraphics[width=1.05\textwidth]{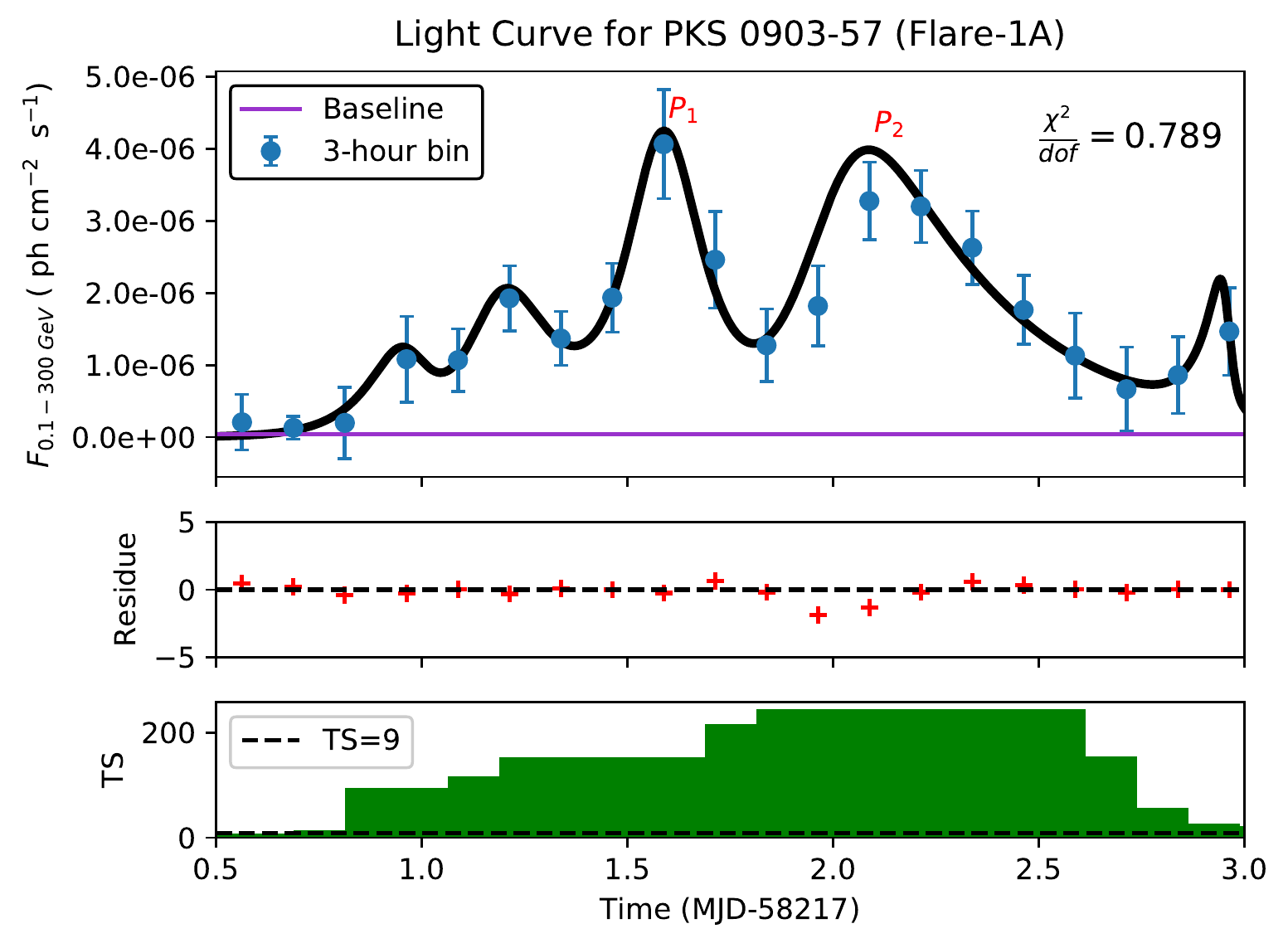}
    \caption{Fitted light curve with \autoref{Eqn:LC_Fitting} of Flare-1A (MJD 58217.5-58220.0)}
     \label{Fig:Flare-1A_LC_Fitting}                    
    \end{minipage} \hfill
    \begin{minipage}{0.49\textwidth}
    \includegraphics[width=1.05\textwidth]{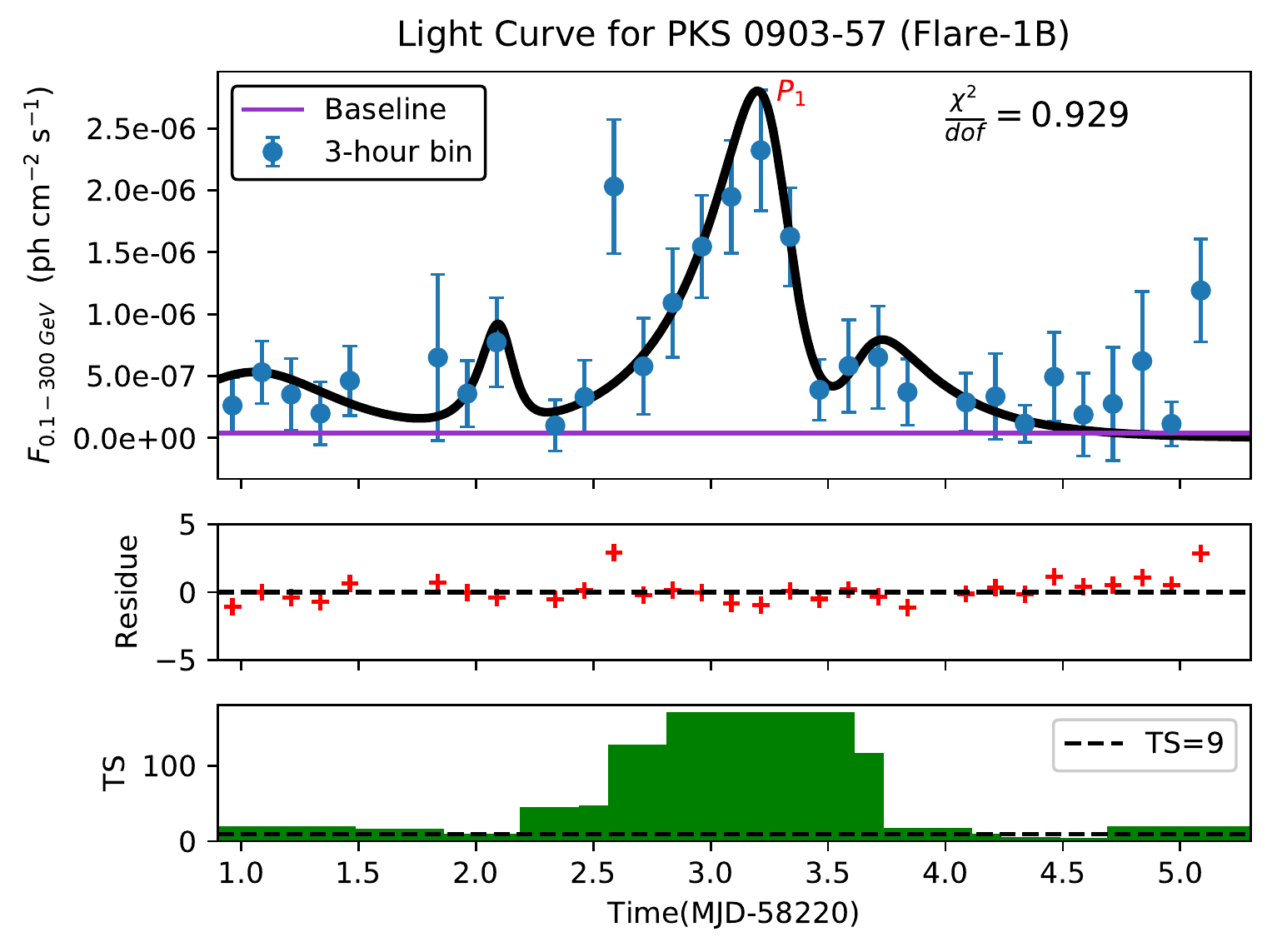}
    \caption{Fitted light curve with \autoref{Eqn:LC_Fitting} of Flare-1B (MJD 58220.9-58225.3)}
    \label{Fig:Flare-1B_LC_Fitting}
    \end{minipage} 
\end{figure}

\begin{figure}
\centering
        \begin{minipage}{0.49\textwidth}
        \includegraphics[width=0.99\textwidth]{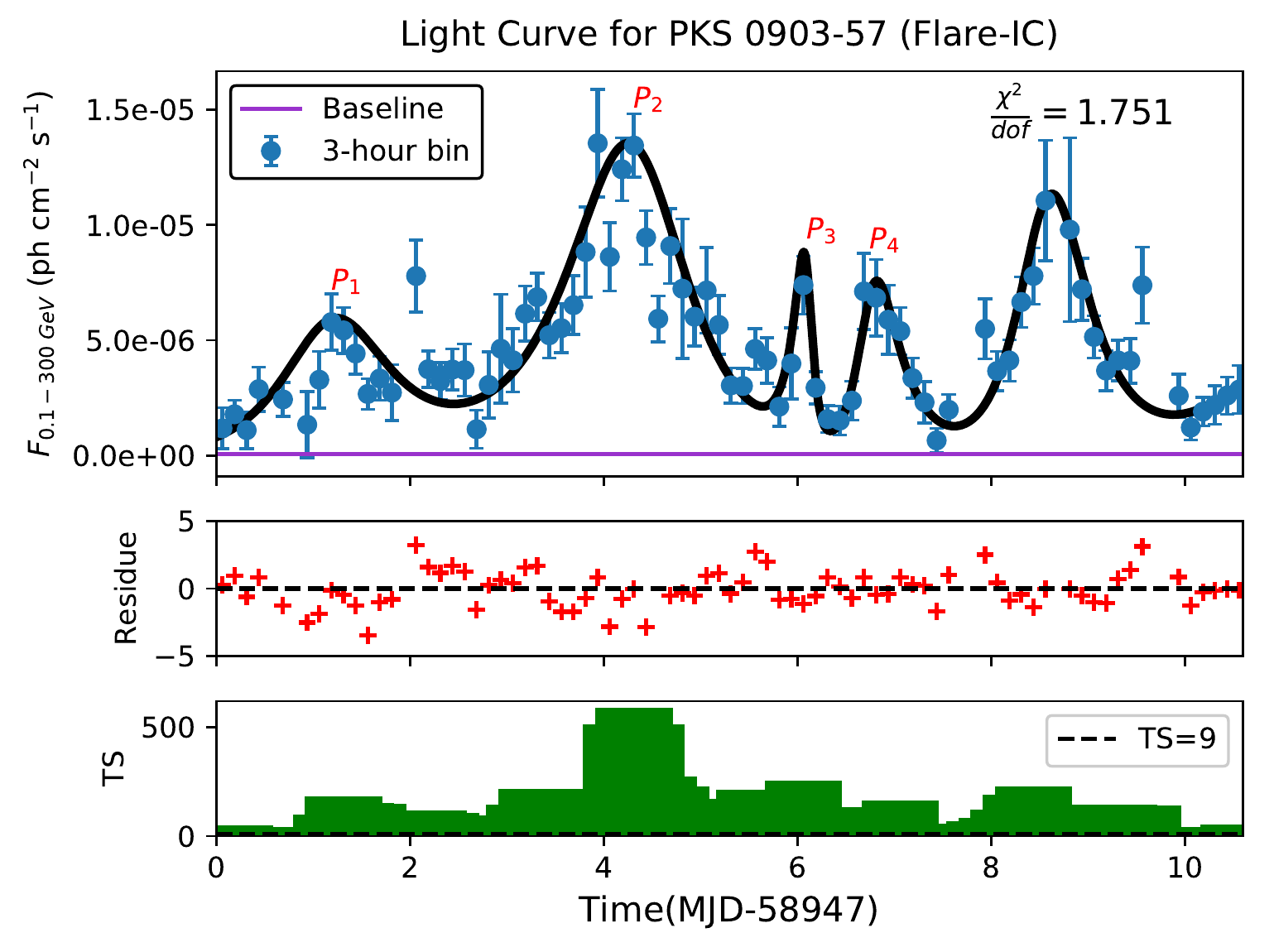}
        \caption{Fitted light curve with \autoref{Eqn:LC_Fitting} of Flare-IC (MJD 58947.0-58957.6)}
        \label{Fig:Flare-IC_LC_Fitting}                            
        \end{minipage}\hfill
       \begin{minipage}{0.49\textwidth}
        \includegraphics[width=0.99\textwidth]{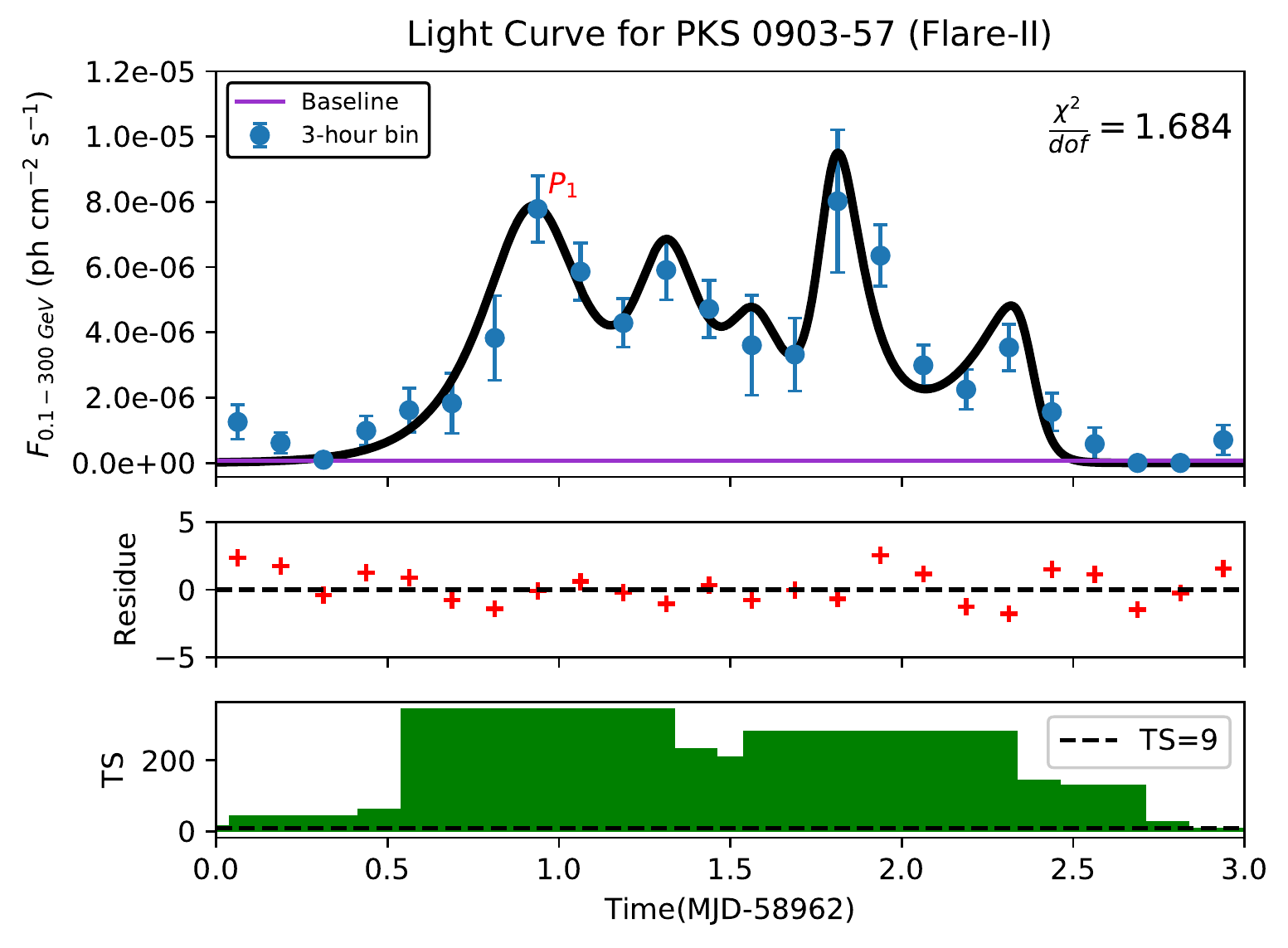}
        \caption{Fitted light curve with \autoref{Eqn:LC_Fitting} of Flare-II (MJD 58962.0-58965.0)}
        \label{Fig:Flare-II_LC_Fitting}
        \end{minipage}

\end{figure}

   We have studied each flare and its activity states or phases (e.g. preflare, flare and postflare) separately as shown in \autoref{Fig:Flare-1_Multi_Bin}, \autoref{Fig:Flare-I_Multi_Bin} \&  \autoref{Fig:Flare-II_Multi_Bin} for Flare-1, Flare-I \& Flare-II respectively. There are several methods to define the different phases of a source. We have discussed the following two methods to define the different phases of a source.
 \begin{itemize}
  \item We have used `Bayesian Blocks' method \citep{scargle2013bayesian} to determine the flaring phases. We have applied this method to `Flare-1' \& `Flare-2' (shown in \autoref{Fig:7D_Whole}, the application of this method on both of the flares has been shown separately in \autoref{Fig:Flare-1_Bayesian_Block} \& \autoref{Fig:Flare-2_Bayesian_Block} respectively). In every case, a segment can be called `Flare' when the flux value is above 5$\sigma$ about the mean flux.

  \item Estimation of each phase's average flux (preflare, flare, etc.) and compare their values. If the average flux of a particular phase is more than 3-4 times of the average flux during preflare or postflare, that particular phase can be defined as `Flare'. We have tabulated the average $\gamma$-ray flux of different phases in \autoref{Tab:Avg_GammaR_Flux}, where we can see the average $\gamma$-ray flux of the flaring phases are 3-4 times higher than the `preflare' or `postflare' states. 
  
 \end{itemize}
We have studied the temporal evolution of each flaring phases separately. Each flaring phases consists of one or more peaks and there are rising and decay time corresponding to each peak and the data points below the detection limit of 3$\sigma$ (TS$<$9) have been rejected for the temporal study. We have fitted the 3-hour binned $\gamma$-ray light curve of each flaring phases with a sum of exponential function. The functional form is given below \citep{Abdo_2010}
	\begin{equation}
	F(t)=2F_{\circ}\Big[exp\Big(\frac{t_{\circ}-t}{T_r}\Big)+exp\Big(\frac{t-t_{\circ}}{T_d}\Big)\Big]^{-1}
	\label{Eqn:LC_Fitting}
	\end{equation}
	where, t$_{\circ}$ is the peak time when the $\gamma$-ray flux is highest within a specific period, F$_{\circ}$ is the flux observed at time t$_{\circ}$ also called as `peak flux', $T_{r}$ is rising time and $T_{d}$ is decay time of the peak. Each of the figures (\autoref{Fig:Flare-1A_LC_Fitting}- \autoref{Fig:Flare-II_LC_Fitting}) consists of three panels; the upper panel shows the $\gamma$-ray light curve fitted with \autoref{Eqn:LC_Fitting}, middle one shows the residual plot and the lower panel shows the TS (Test-Statistics) plot of the data points. A horizontal dark-orchid line in the upper panel has been shown in \autoref{Fig:Flare-1A_LC_Fitting} to \autoref{Fig:Flare-II_LC_Fitting}, which is the baseline flux. In a few cases the light curve fittings seem to be over-fitted due to the following reasons: if the peak that we defined during a flaring phase, consists of a single data point then it is very difficult to fit that peak. Also, if the points include large error bars then the fit may be over-fitted. Both of which are true in our case. We have showed a residual plot corresponding to each fitted light curve where we have plotted time vs residue to show the quality of fitting. The residue is defined as the ratio of the difference between model and observed flux to the flux error. It can be seen that the residual calculated for each data point is confined within $\pm$3$\sigma$ confidence level. There are only very few points which are out of this zone. In case of the TS plot, we have drawn a baseline of TS=9 to show that the TS value of the  data points are much higher than 9 i.e.  they are detected with much higher confidence level, so their detection is highly significant.

\section{Description of Flares}
\label{Sec:Sec5_Flare_Descriptions}
In this section, we have discussed about the flares in details.

	\subsection{Flare-1}
	\autoref{Fig:Flare-1_Multi_Bin} shows the $\gamma$-ray light curve of Flare-1 in time bins of 1-day, 12-hour, 6-hour and 3-hour corresponding to the flaring activity of PKS 0903-57 between MJD 58216.5 to 58230. Only one sub-structure has been detected. Also phases, which can be seen in \autoref{Fig:Flare-1_Multi_Bin}, are prominently visible in 12-hour, 6-hour and 3-hour time bin. The flaring activity of Flare-1 can be divided into two phases: Flare-1A and Flare-1B. The $\gamma$-ray flux before Flare-1A and after Flare-1B are too low for analysis, so we have not considered any of the two regions as Preflare or Postflare. This activity was observed by \textit{Fermi}-LAT \citep{2018ATel11644....1C} in May 2018.\par
     We have used `Bayesian Block' method to detect different activity phases as shown in \autoref{Fig:Flare-1_Bayesian_Block}. \par
	Flare-1A was observed between MJD 58217.5 to 58220.0, persisted for $\sim$3 days. \autoref{Fig:Flare-1A_LC_Fitting} shows the temporal evolution of $\gamma$-ray flux during Flare-1A, where we can see two major peaks: P$_1$ and P$_2$ around MJD 58218.59 and MJD 58219.08 with flux 4.06$\pm$0.75 \& 3.27$\pm$0.53 respectively. The average $\gamma$-ray flux during this flare is 1.8$\pm$0.5. Similarly, \autoref{Fig:Flare-1B_LC_Fitting} shows the temporal evolution of $\gamma$-ray flux during Flare-1B which was observed between MJD 58220.9 to 58225.3, where we can see a single peak, P$_1$ at MJD 58223.28 with flux 2.32$\pm$0.48. And the average flux during this period is 0.6$\pm$0.4.\par
	The flares have been fitted with \autoref{Eqn:LC_Fitting}. The decay time and the rising time of the peaks are tabulated in \autoref{Tab:Flare-1A_RD_Time} and \autoref{Tab:Flare-1B_RD_Time} for Flare-1A \& Flare-1B respectively.

	\subsection{Flare-2}
\begin{figure}[h]
\centering
\includegraphics[width=0.94\textwidth]{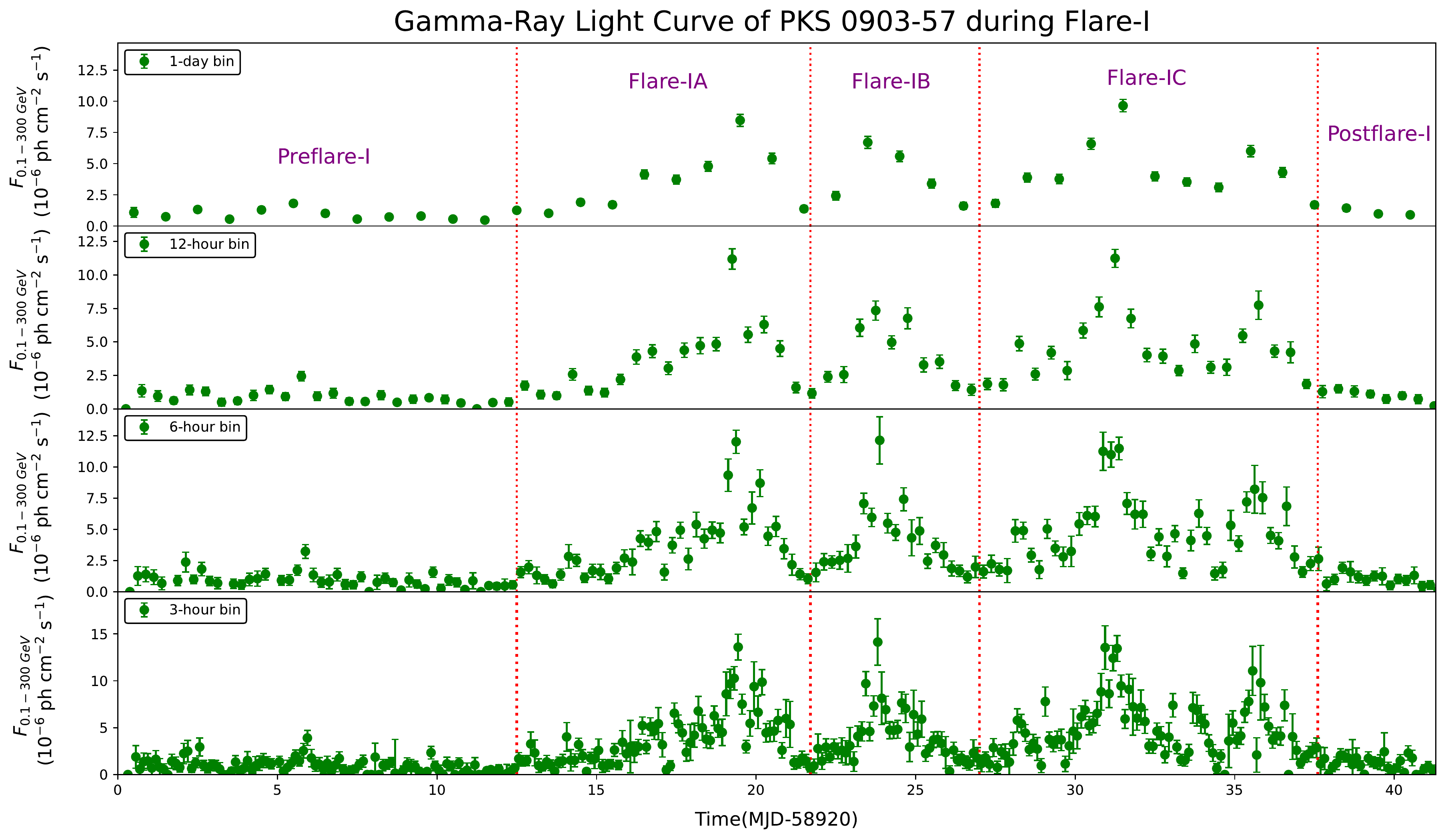}
\caption{ $\gamma$-ray light curve of PKS 0903-57 during Flare-I (MJD 58920.0-58961.3), sub-structure of Flare-2 shown in \autoref{Fig:Flare-I_&_II_Combo}. Five phases of Flare-I: Preflare-I, Flare-IA, Flare-IB, Flare-IC \& Postflare-I.}
\label{Fig:Flare-I_Multi_Bin}
\end{figure}

\begin{figure}
\centering
\includegraphics[width=0.91\textwidth]{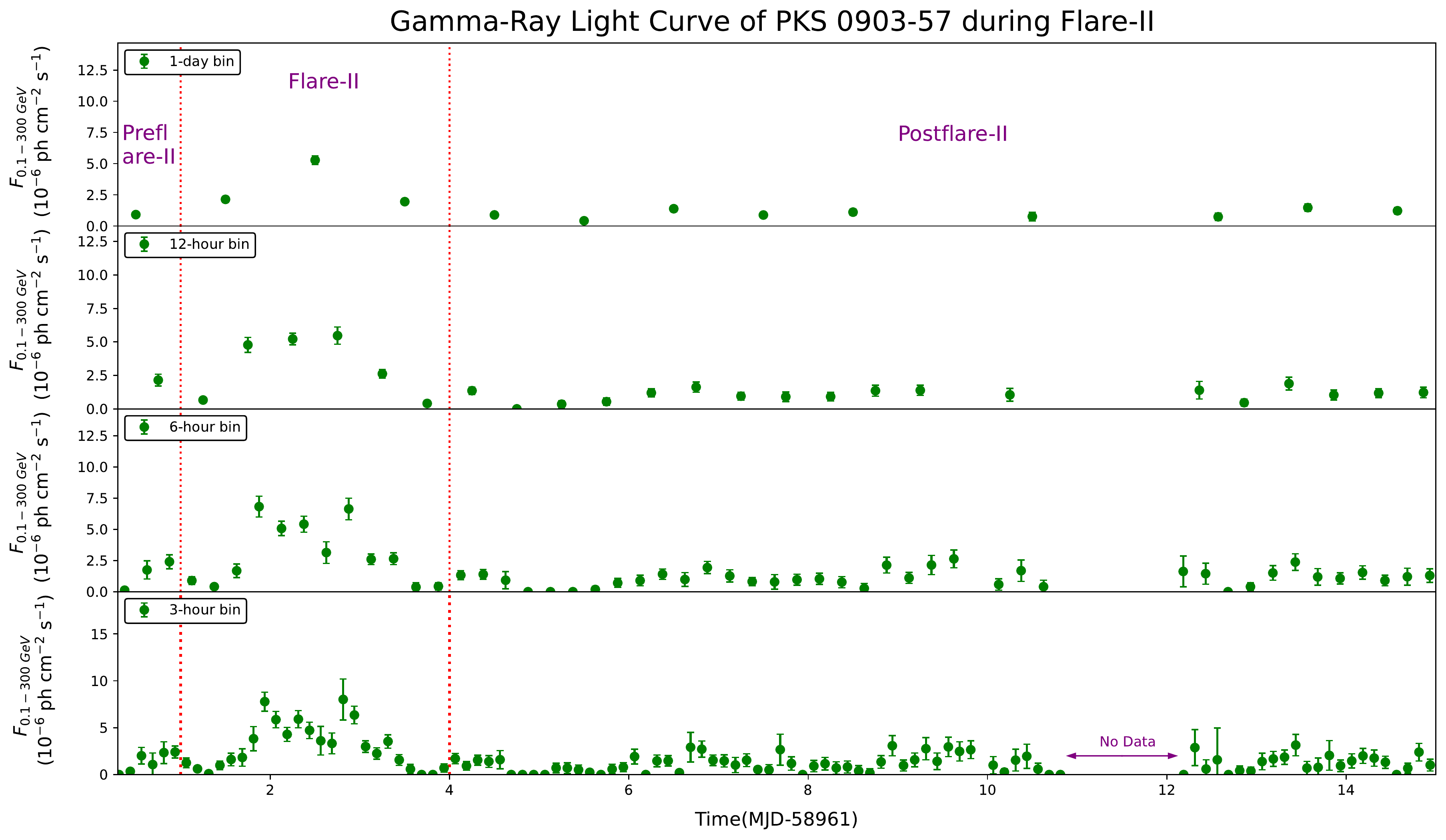}
\caption{$\gamma$-ray light curve of PKS 0903-57 during Flare-II (MJD 58961.3-58976.0), sub-structure of Flare-2 shown in \autoref{Fig:Flare-I_&_II_Combo}. Three phases of Flare-II: Preflare-II, Flare-II \& Postflare-II.}
\label{Fig:Flare-II_Multi_Bin}
\end{figure}

	From MJD 58920 to MJD 58976, another flaring activity, Flare-2, of PKS 0903-57 has been observed as shown in \autoref{Fig:7D_Whole}. In shorter time binning,  we have found that Flare-2 has two sub-structures, shown in \autoref{Fig:Flare-I_&_II_Combo}, denoted by Flare-I  \& Flare-II.  In a shorter time bin, the phases of Flare-I \& Flare-II are more prominent which are shown in \autoref{Fig:Flare-I_Multi_Bin} \& \autoref{Fig:Flare-II_Multi_Bin} respectively.
	 
	This flaring activity of PKS 0903-57 was reported (between end of March 2020 to April 2020) by \textit{Fermi}-LAT \citep{2020ATel13599....1M} \citep{2020ATel13604....1B}, AGILE \citep{2020ATel13602....1L}, HESS \citep{2020ATel13632....1W} and DAMPE \citep{2020ATel13643....1D}. This was reported as the brightest flare ever detected by \textit{Fermi}-LAT from this source.\par
	In \autoref{Fig:Flare-2_Bayesian_Block}, we have shown the application of the `Bayesian Block' method to detect different activity phases of Flare-2.
	
	Flare-I has five distinct phases: Preflare-I, Flare-IA, Flare-IB, Flare-IC \& Postflare-I, shown in \autoref{Fig:Flare-I_Multi_Bin}. Each region is prominently visible in the $\gamma$-ray light curve in shorter time bins.
	
	Preflare-I has been observed from MJD 58920 to 58932.5, over 12 days where the $\gamma$-ray flux is very low; after this phase, a rise in the $\gamma$-ray flux has been observed. The average flux during this period is 0.9$\pm$0.6.
\par

\begin{figure}
\centering
        \begin{minipage}{0.49\textwidth}
        \includegraphics[width=0.99\textwidth]{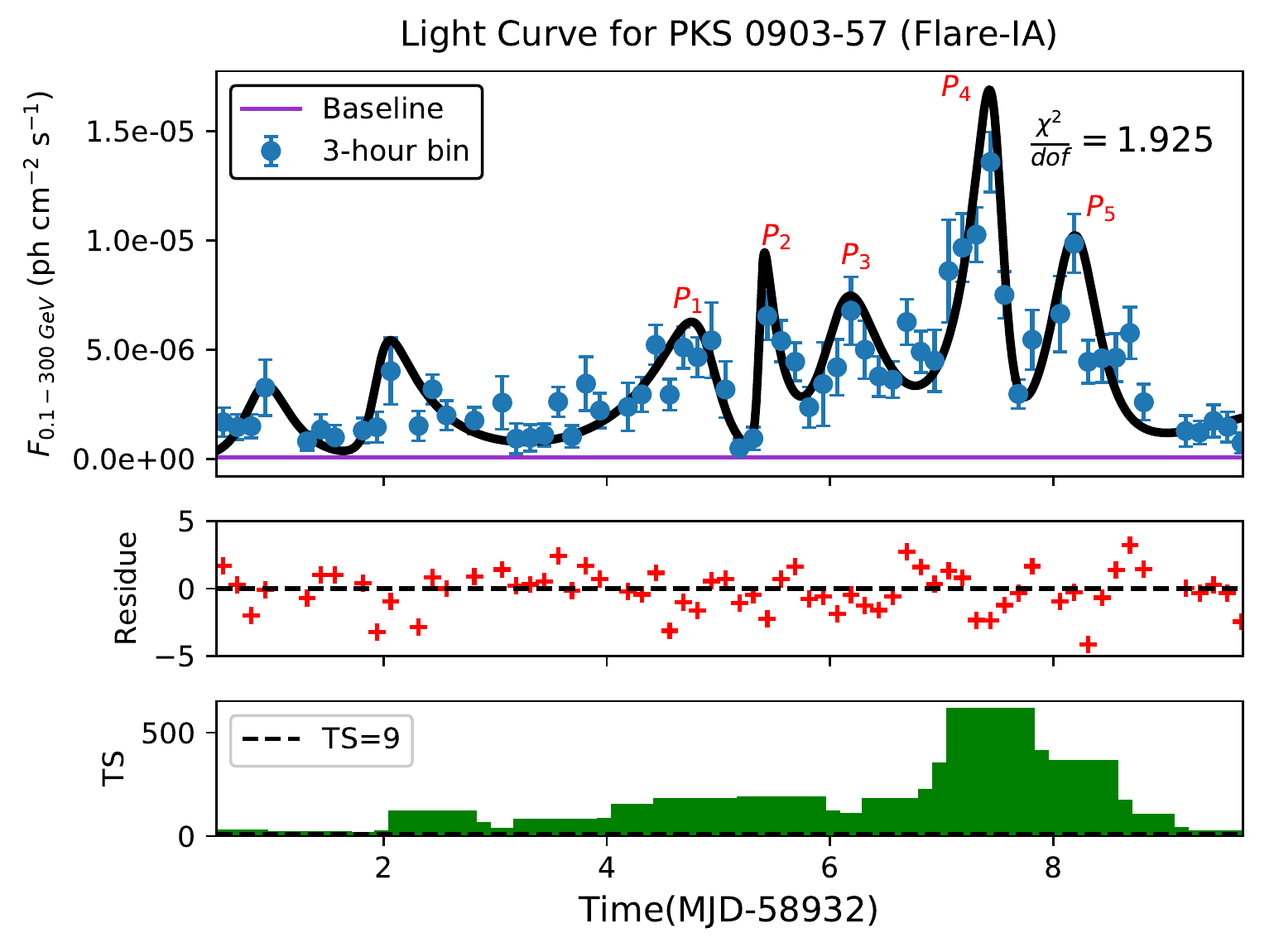}
        \caption{Fitted light curve with \autoref{Eqn:LC_Fitting} of Flare-IA (MJD 58932.5-58941.7)}
        \label{Fig:Flare-IA_LC_Fitting}
        \end{minipage} \hfill 
        \begin{minipage}{0.49\textwidth}
        \includegraphics[width=0.99\textwidth]{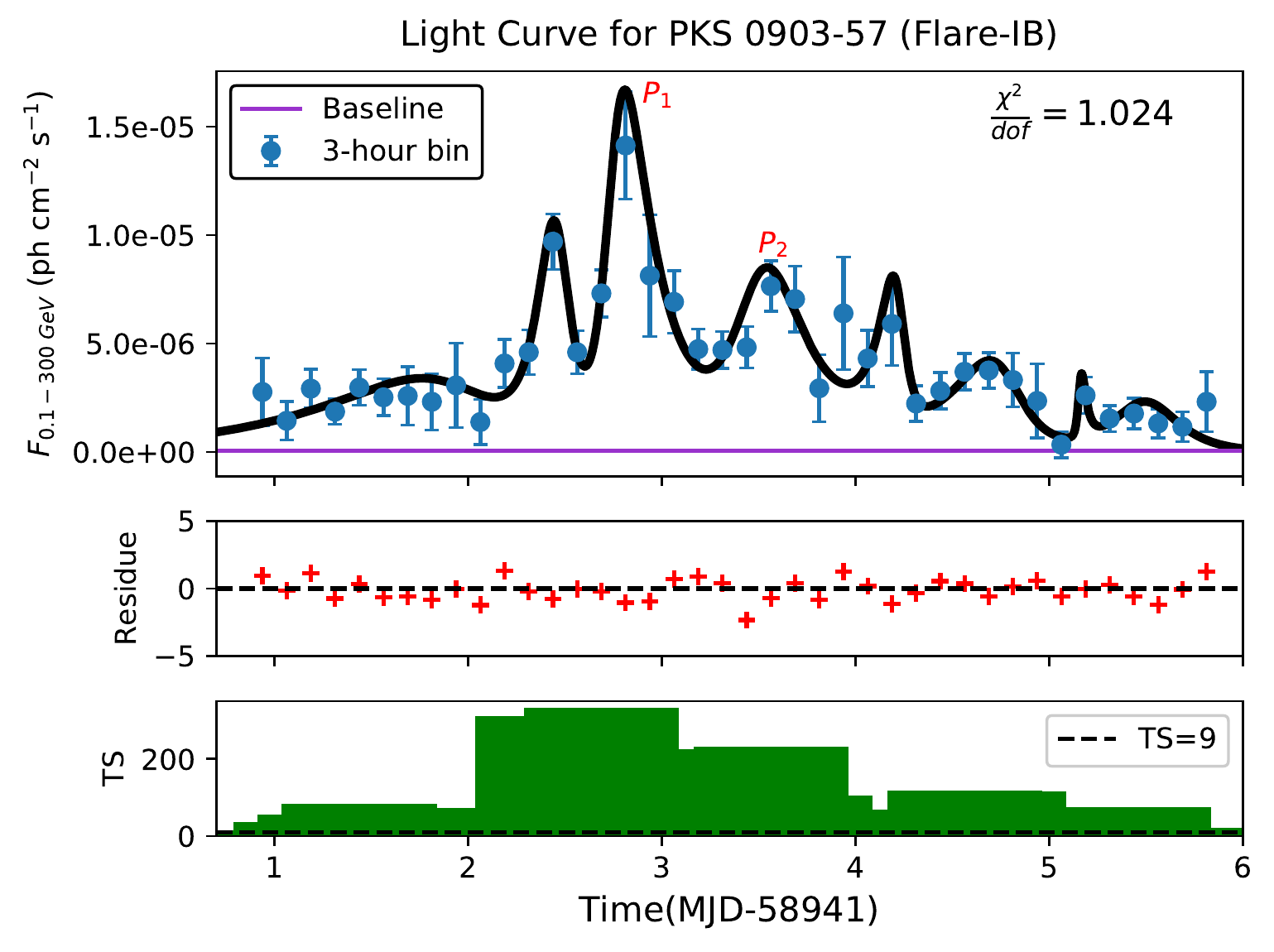}
        \caption{Fitted light curve with \autoref{Eqn:LC_Fitting} of Flare-IB (MJD 58941.7-58947.0)}
        \label{Fig:Flare-IB_LC_Fitting}                            
        \end{minipage}
\end{figure}

Preflare-I is followed by three flaring phases. These three flaring segments are Flare-IA, Flare-IB and Flare-IC respectively. Flare-IA was observed from MJD 58932.5 to 58941.7, which persisted almost for 9 days. In \autoref{Fig:Flare-IA_LC_Fitting}, we can see five peaks P$_1$, P$_2$, P$_3$, P$_4$ and P$_5$ at MJD 58936.90, 58937.38, 58938.10, 58939.50 and 58940.22 respectively and the corresponding fluxes are 5.10$\pm$0.95, 6.54$\pm$1.08, 6.77$\pm$1.55, 13.59$\pm$1.37, 9.86$\pm$1.34 respectively. In \autoref{Fig:Flare-IA_LC_Fitting} we have shown, fitted light curve with  \autoref{Eqn:LC_Fitting} and the decay and rising time are reported in \autoref{Tab:Flare-IA_RD_Time}. The average $\gamma$-ray flux during this period is 3.6$\pm$1.0.
\par
	Flare-IB has been observed between MJD 58941.7 to 58947.0. The temporal evolution of Flare-IB has been shown in \autoref{Fig:Flare-IB_LC_Fitting} with two peaks. The highest peak occurred at MJD 58943.76 with the flux 14.13$\pm$2.46, denoted as P$_1$ and the second peak, P$_2$ is observed at MJD 58944.50 with flux 7.64$\pm$1.16. The decay and rise time corresponding to P$_1$ and P$_2$ are mentioned in \autoref{Tab:Flare-IB_RD_Time}.
	 The average $\gamma$-ray flux during this period is 3.9$\pm$1.2.
	
 Flare-IC has been observed between MJD 58947.0 to 58957.6. This phase persisted almost for 10 days. \autoref{Fig:Flare-IC_LC_Fitting} shows four peaks P$_1$,P$_2$,P$_3$ and P$_4$ at MJD 58948.19, 58951.31, 58953.10 and 58953.75 respectively and corresponding $\gamma$-ray fluxes are 5.78$\pm$1.22, 13.44$\pm$1.37, 7.39$\pm$1.25 and 6.84$\pm$1.65 respectively. The decay and rise time are tabulated in the \autoref{Tab:Flare-IC_RD_Time}. The average $\gamma$-ray flux in this phase is 4.6$\pm$1.2. 
\par
A postflare phase (Postflare-I) is observed between MJD 58957.6 to  58961.3 with the average flux 1.1$\pm$0.7. 
Just after the Postflare-I, a rise in the $\gamma$-ray flux is seen between MJD 58961.3 to 58976.0. This state is defined as Flare-II and the corresponding $\gamma$-ray light curve is shown in \autoref{Fig:Flare-II_Multi_Bin}. The $\gamma$-ray light curves in 1-day, 12-hour, 6-hour and 3-hour time bin have been shown here. 
	This flare consists of three phases: Preflare-II, Flare-II and Postflare-II. The preflare phase lasted for only 1.3 days (MJD 58961.3 to 58962.0); during this period, the average $\gamma$-ray flux is found to be 1.2$\pm$0.6. Preflare-II is followed by a flaring phase (MJD 58962.0 to 58965.0) which is also very short ($\sim$2 days). In \autoref{Fig:Flare-II_LC_Fitting}, we have shown the flare in 3-hour time bin, with a single peak $P_1$ observed at MJD 58962.94 with flux value 7.78$\pm$1.01. The decay and rise time are mentioned in  \autoref{Tab:Flare-II_RD_Time}. The average $\gamma$-ray flux during this period is 4.6$\pm$1.0.
	After the flaring phase, Postflare-II is observed between MJD 58965.0 to 58976.0 with an average flux 1.1$\pm$0.7.

\subsection{Variability Time}
Variability time is a measure of the time scale of variation in flux during flares. 

\begin{equation}
F(t_2)=F(t_1)2^\frac{t_2-t_1}{T_{d/h}}
\end{equation}
where, F(t$_1$) and F(t$_2$) are the fluxes measured at two consecutive time instants t$_1$ and t$_2$ respectively, $T_{d/h}$ denotes flux doubling or halving time which is tabulated in \autoref{Tab:Flux_DH_Time} (`positive' and `negative' value of $T_{d/h}$  in the table denotes doubling and halving time respectively). Two criteria have been kept in mind during the scanning of the $\gamma$-ray light curve \citep{Prince_2017} :
\begin{itemize}
\item Only those consecutive time instants will be considered which have TS$>$25 ($>$5$\sigma$ detection; \citealt{1996ApJ...461..396M}).
\item The flux ratio between these two time instants should be greater than two (rising part) or less than half (decaying part). 
\end{itemize}
There are several consecutive time instants with flux ratio more than two or less than half but the TS value of those observations are less than 25, we have not included these cases.

In our 12 years $\gamma$-ray light curve study,  the shortest $\gamma$-ray flux doubling/halving time ($T_{d/h}$) is found to be 1.7$\pm$0.9 hour (mentioned in \autoref{Tab:Flux_DH_Time}, during Flare-I  for MJD 58935.688 \& 58935.813).

\section{Gamma- Ray Spectral Energy Distribution of different flaring phases}
\label{Sec:Sec6_GammaR_SED}
We have fitted different phases (e.g. preflare, flare, postflare) of the activity periods with four different spectral models. The details about the models are the following:
\begin{enumerate}
\item PowerLaw (PL) :\\
The functional form of the powerlaw is the following,
	\begin{equation}
	\frac{dN}{dE}=N_{\circ}\Bigg(\frac{E}{E_{\circ}}\Bigg)^{-\Gamma} 
	\end{equation}
where, N$_{\circ}$ is the prefactor, $\Gamma$ is the powerlaw index and E$_{\circ}$ is the scaling factor or pivot energy. We have kept a fixed value of E$_{\circ}$ which is 1155.4126 MeV \citep{Abdollahi_2020} for all the $\gamma$-ray SEDs of this source.

\item LogParabola (LP) : \\
The functional form of the logparabola is the following, 
	\begin{equation}
	\frac{dN}{dE}=N_{\circ}\Bigg(\frac{E}{E_{\circ}}\Bigg)^{-(\alpha+\beta log(E/E_{\circ}))} 
	\end{equation}
where, N$_{\circ}$ is the prefactor, $\alpha$ is photon index, $\beta$ is curvature index. Scaling factor (E$_{\circ}$) is fixed to 1155.4126 MeV similar to the powerlaw function.

\item BrokenPowerLaw (BPL) :\\
The functional form of the brokenpowerlaw is the following,
    \begin{equation}
 		\frac{dN}{dE} =N_{\circ} 
						\begin{cases}
   					    \Big(\frac{E}{E_b}\Big)^{-\Gamma_1},& \text{for} \ E<E_b  \\
  						 \Big(\frac{E}{E_b}\Big)^{-\Gamma_2} ,& \text{otherwise}
						\end{cases}
	\end{equation}
where,  N$_{\circ}$ is prefactor, $\Gamma_1$ and $\Gamma_2$ are spectral indices , E$_b$ is the break energy.	
	
\item PowerLaw with Exponential Cutoff	(PLEC) : \\
The functional form of the PLEC is the following,
  \begin{equation}
  \frac{dN}{dE}=N_{\circ}\Big(\frac{E}{E_{\circ}}\Big)^{-\Gamma_{PLEC}}exp\Big(-\Big(\frac{E}{E_c}\Big)\Big)
  \end{equation}
  
  where, N$_{\circ}$ is prefactor, $\Gamma_{PLEC}$ is the PLEC index, E$_{\circ}$ is pivot energy which is fixed at 1155.4126 MeV similar to powerlaw and $E_c$ is cutoff energy.
\end{enumerate} 

  We have used the maximum likelihood fitting to determine the best-fit model.
   In \autoref{Fig:Flare-1_GammaR_SED}, we have shown \textit{Fermi}-LAT SEDs of Flare-1 for its two sub-structures: Flare-1A \& Flare-1B. Both the SEDs have been fitted with four spectral models: PowerLaw (PL), LogParabola(LP), BrokenPowerLaw(BPL) \& PowerLaw with Exponential Cutoff (PLEC). Black, red, magenta \& blue color have been used to denote the fitting of the spectral points with PL, LP, BPL \& PLEC respectively.
   
  \autoref{Tab:Flare-1_GammaR_SED_Param} contains all the parameter values that have been used to fit the \textit{Fermi}-LAT $\gamma$-ray spectral points of Flare-1A \& Flare-1B with the above mentioned spectral models. In this table, we have mentioned the fitted flux, spectral indices, TS and -log(Likelihood) values.
  
 We have also calculated the $\triangle$log(Likelihood) value \citep{Britto_2016} which is defined as  $\triangle$log(Likelihood)=(-log(Likelihood)$_{LP/BPL/PLEC}$)-(-log(Likelihood)$_{PL}$). 
 
\begin{figure}[h]
\centering
\includegraphics[width=0.49\textwidth]{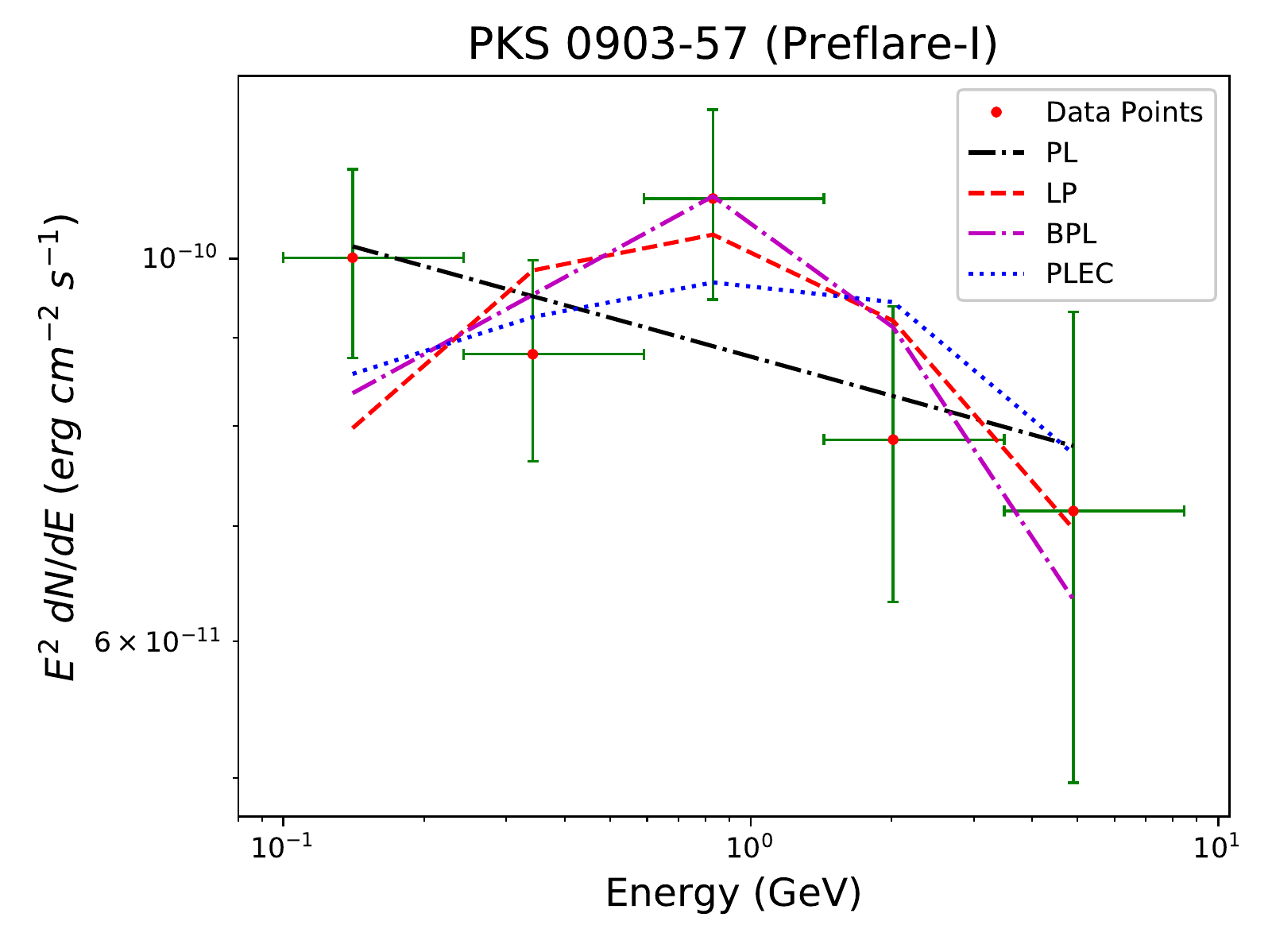}
\includegraphics[width=0.49\textwidth]{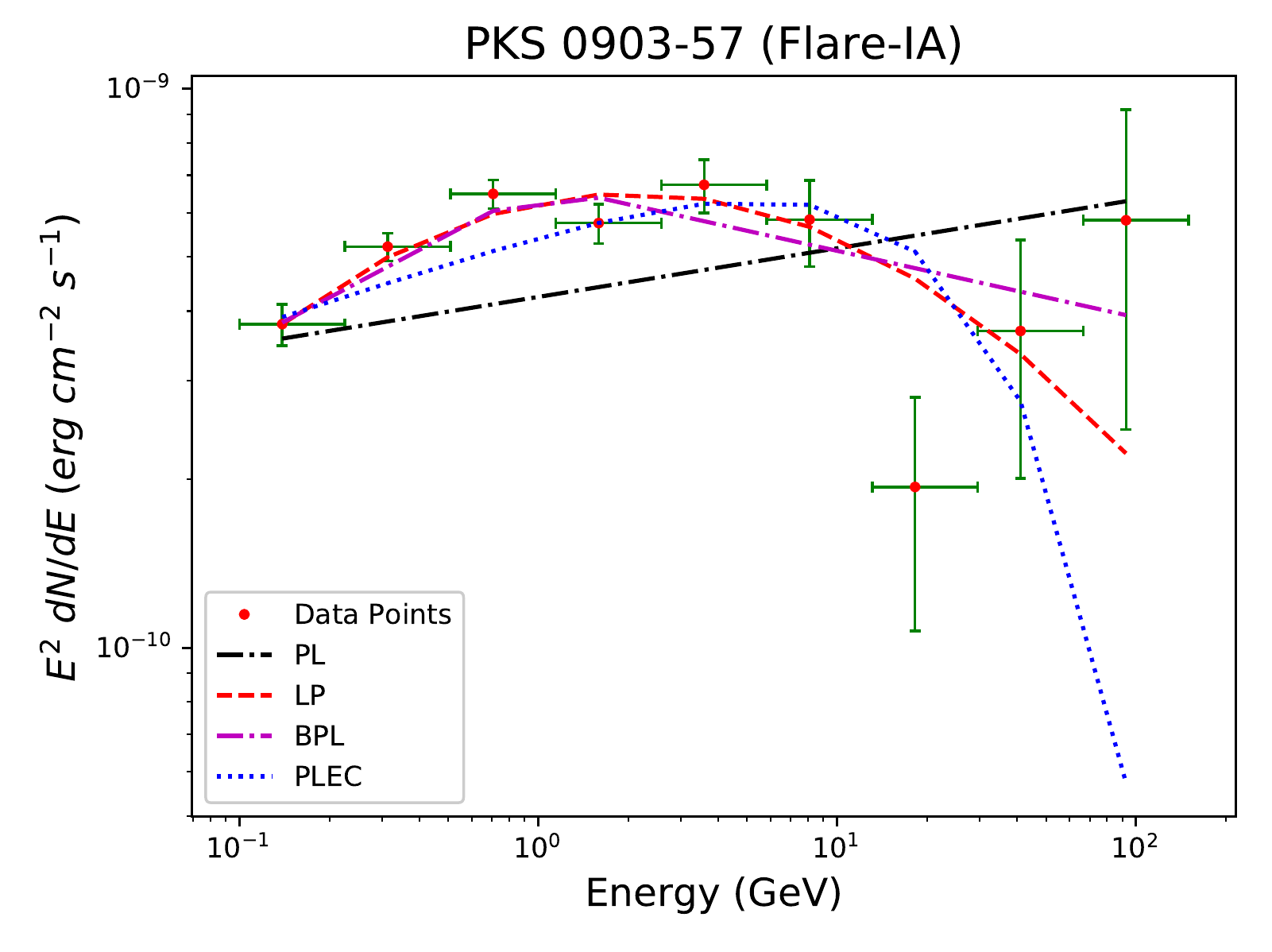}
\includegraphics[width=0.49\textwidth]{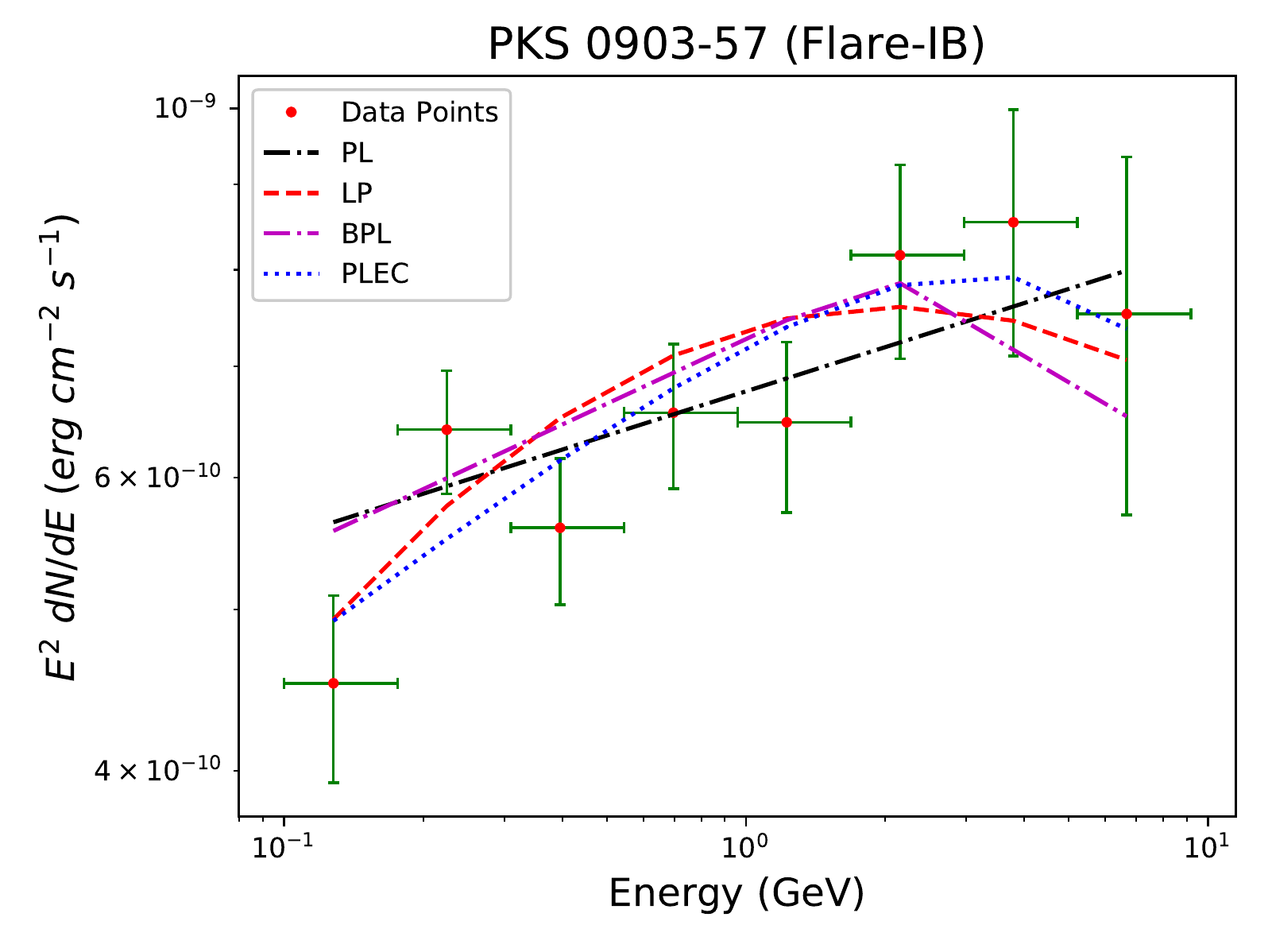}
\includegraphics[width=0.49\textwidth]{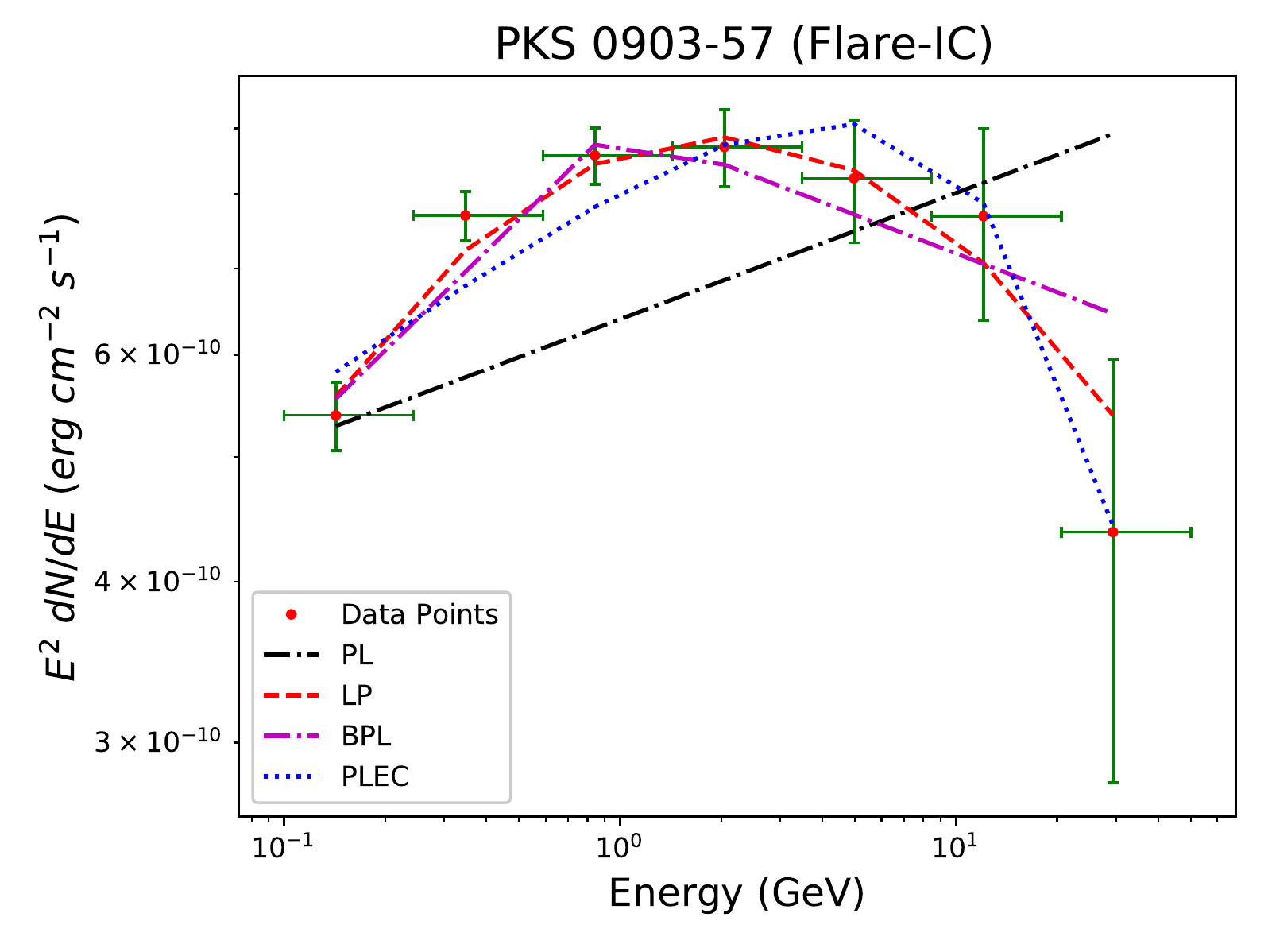}
\includegraphics[width=0.49\textwidth]{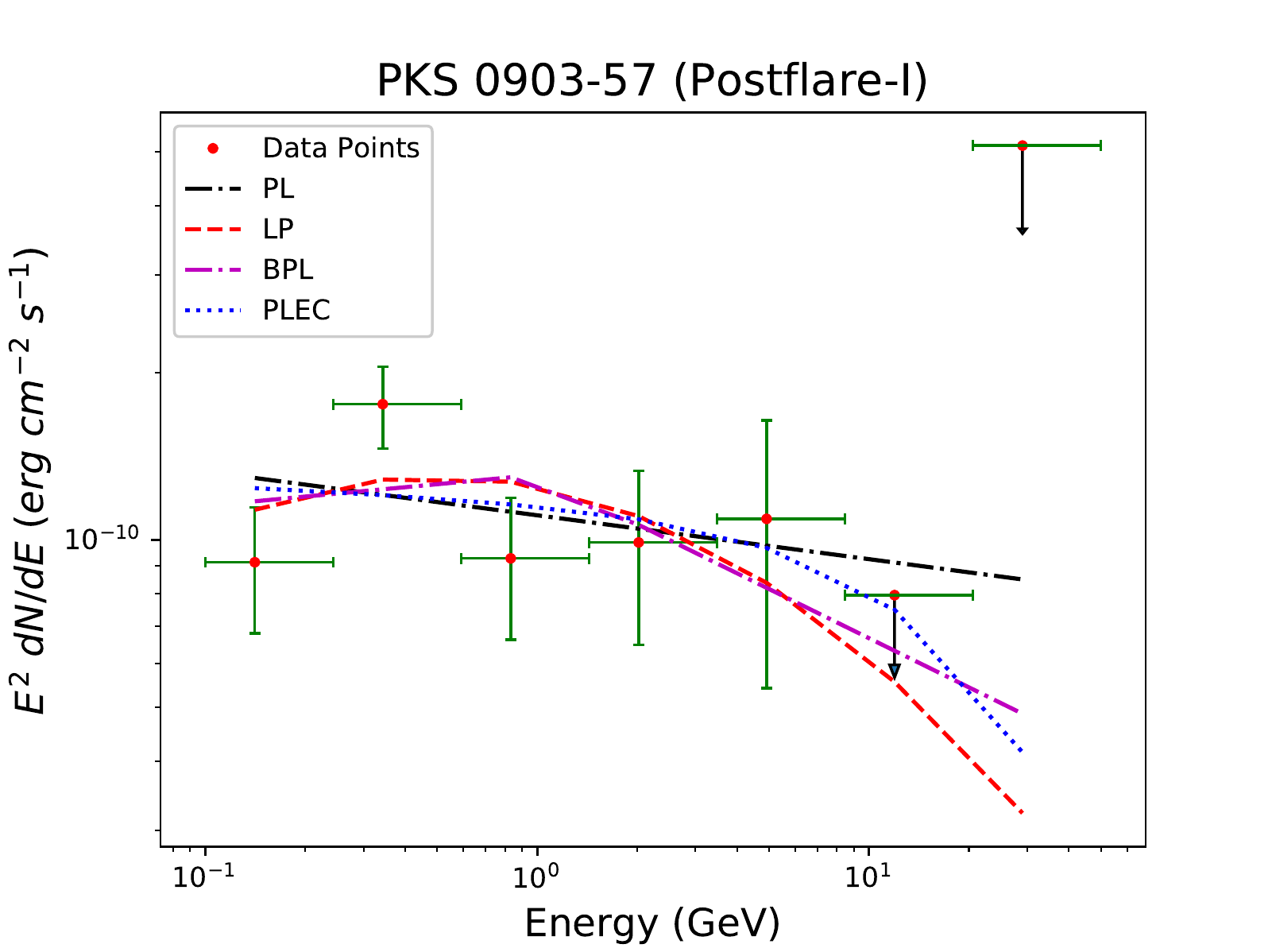}
\caption{$\gamma$-ray SEDs of five activity phases of Flare-I as shown in \autoref{Fig:Flare-I_Multi_Bin}. SEDs have been fitted with four spectral models mentioned earlier.}
\label{Fig:Flare-I_GammaR_SED}
\end{figure}

    In \autoref{Fig:Flare-I_GammaR_SED}, we have shown \textit{Fermi}-LAT $\gamma$-ray SEDs of the five phases of Flare-I. Similarly, all the SEDs have been fitted with the same four spectral models and the fitted parameter values are tabulated in \autoref{Tab:Flare-I_GammaR_SED_Param}.
 
\begin{figure}
\centering
\includegraphics[width=0.49\textwidth]{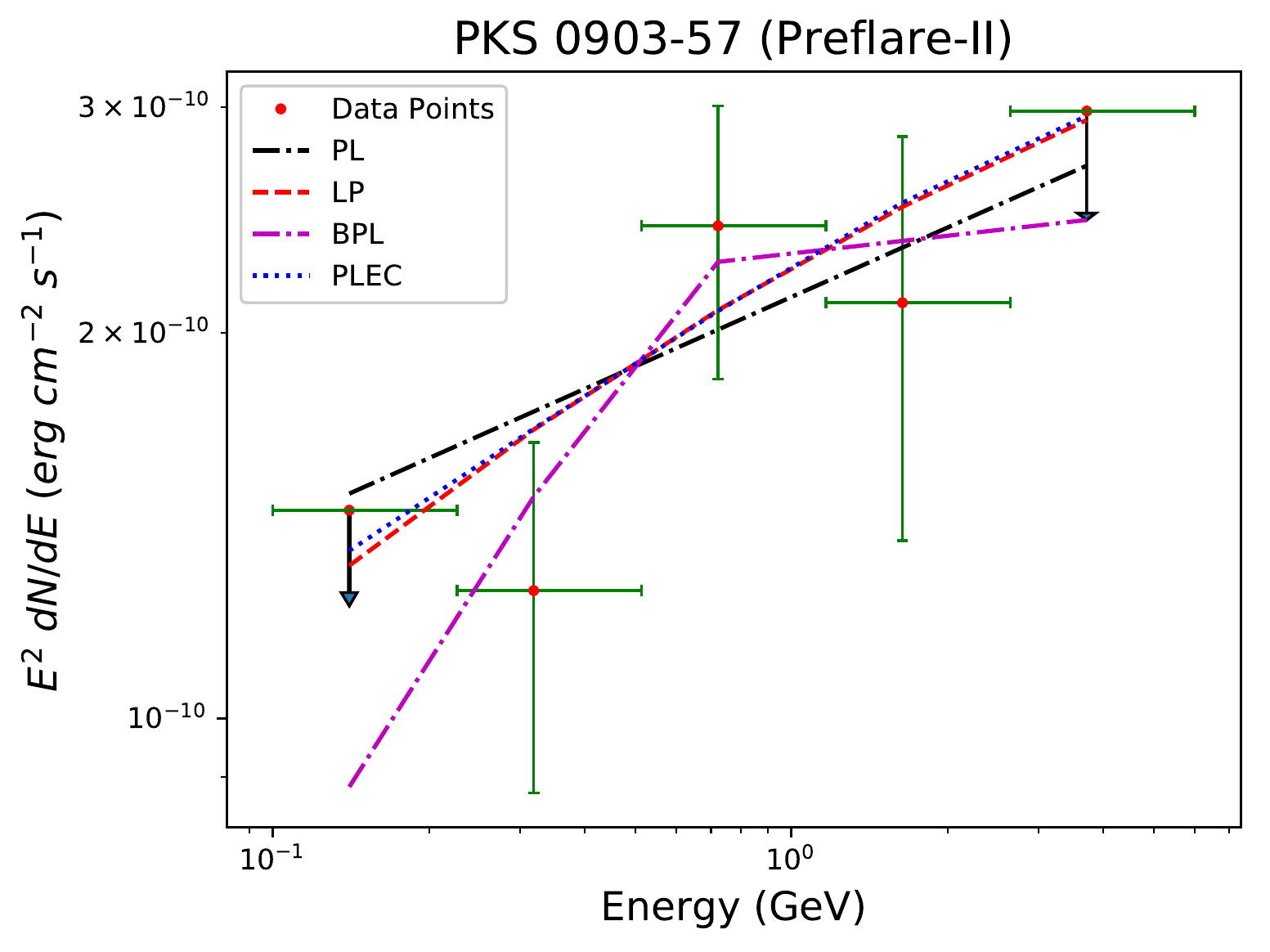}    				\includegraphics[width=0.49\textwidth]{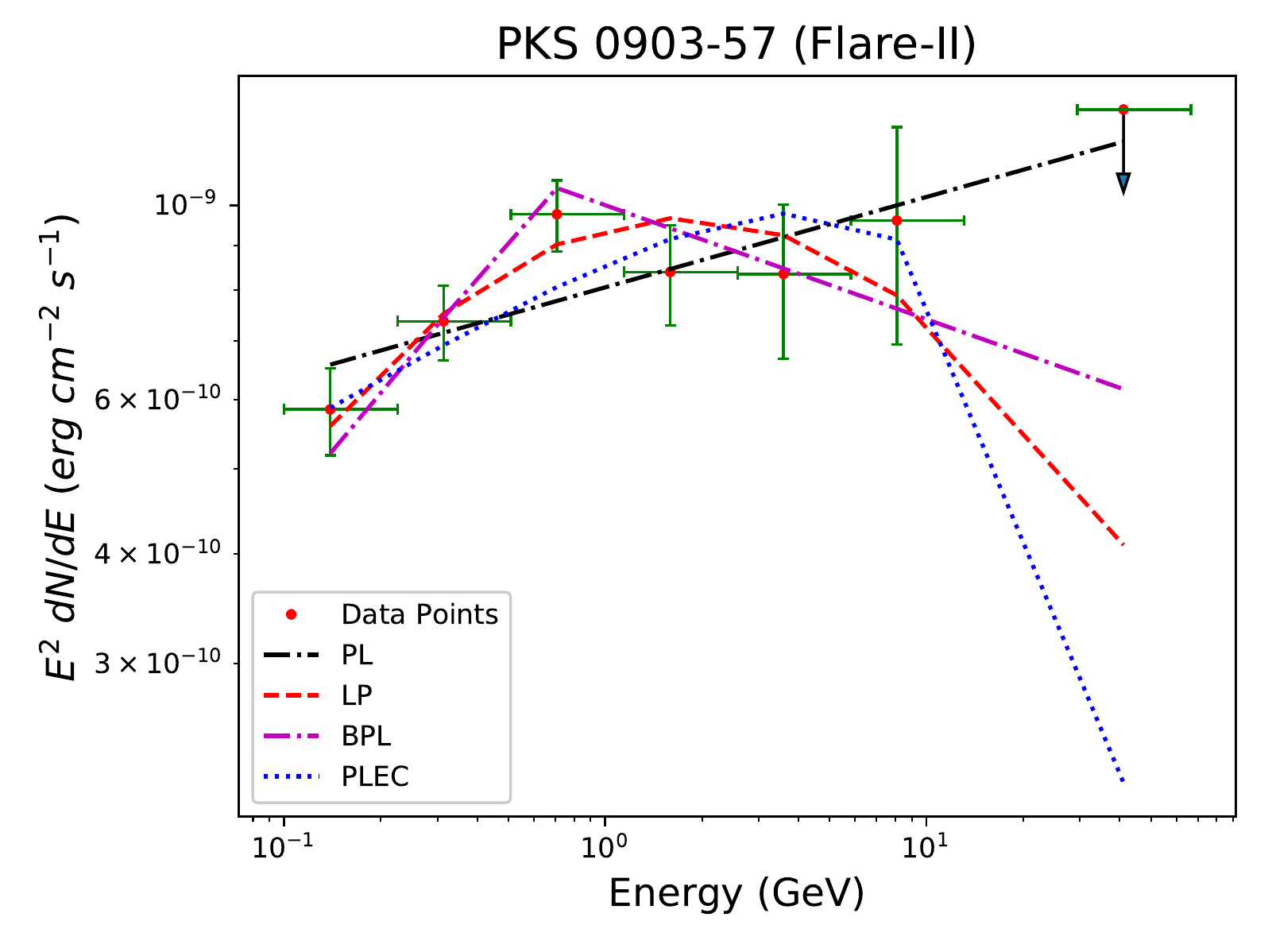} 
\includegraphics[width=0.49\textwidth]{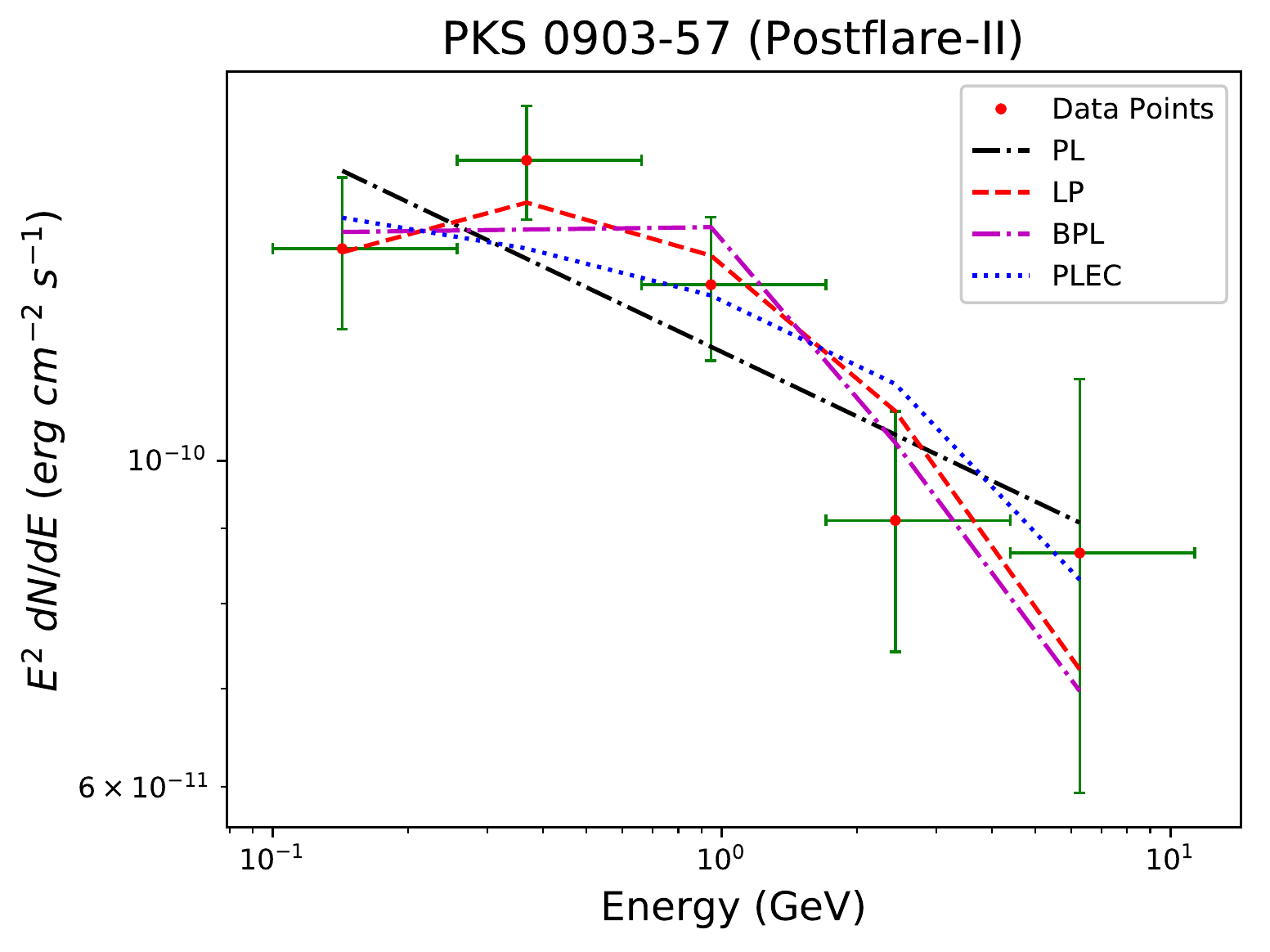}
\caption{$\gamma$-ray SEDs of three activity phases of Flare-II as shown in \autoref{Fig:Flare-II_Multi_Bin}. SEDs have been fitted with the four spectral models mentioned earlier.}
\label{Fig:Flare-II_gammaR_SED}
\end{figure}

In \autoref{Fig:Flare-II_gammaR_SED}, SEDs of three activity phases of Flare-II have been shown, which are fitted with the same four spectral models i.e. PL, LP, BPL \& PLEC and \autoref{Tab:Flare-II_GammaR_SED_param} contains all the fitted parameter values.
 
In case of Flare-1, as the source transits from Flare-1A ($\Gamma$=1.98$\pm$0.05) to Flare-1B ($\Gamma$=1.93$\pm$0.07), the $\gamma$-ray spectral index remains almost constant. 
 
 Flare-I shows spectral hardening when the source transits from Preflare-I ($\Gamma$=2.08$\pm$0.06) to Flare-IA ($\Gamma$=1.91$\pm$0.02) which can be seen from \autoref{Tab:Flare-I_GammaR_SED_Param}. However, during the transition from Flare-IA ($\Gamma$=1.91$\pm$0.02) to Flare-IB ($\Gamma$=1.94$\pm$0.03) \& Flare-IB ($\Gamma$=1.94$\pm$0.03) to Flare-IC ($\Gamma$=1.90$\pm$0.02), the spectral index remains almost constant. The spectrum softens when the source transits from Flare-IC ($\Gamma$=1.90$\pm$0.02) to Postflare-I ($\Gamma$=2.08$\pm$0.10). For Flare-II, the spectrum softens as the source transits from Preflare-II ($\Gamma$=1.80$\pm$0.10) to Flare-II ($\Gamma$=1.92$\pm$0.03).
\par
From the above $\gamma$-ray SED analysis of the source PKS 0903-57, we can see that the $\gamma$-ray spectrum may harden or soften or remain almost unchanged during transition from one phase to another.
 Earlier, \citealt{Das_2020} found spectral hardening as an important feature of the source 3C 454.3. However, we observed all the possibilities for the source PKS 0503-57, in some cases we saw \enquote{brighter-when-harder}, in some cases \enquote{brighter-when-softer} scenario, and in some other cases spectral index remains almost unchanged. 

\par

From the maximum likelihood analysis using different spectral models during the different activity phases, we find that BPL is the best-fit model for Flare-1 \& Flare-II, whereas LP is the best-fit model for Flare-I. We have multi-wavelength data for only these four phases: Flare-1B, Flare-IA, Flare-IB \& Flare-IC. For three of them (Flare-IA, Flare-IB \& Flare-IC) LP is the best-fit model.
 However, in case of Flare-1B the $\Delta$ log(Likelihood) values are very close to each other (see \autoref{Tab:Flare-1_GammaR_SED_Param}) for LP and BPL, and hence both models are preferred.
Therefore, we have used LP model to fit the multi-wavelength SEDs of all the four phases in this work. 

\section{Multi-Wavelength Study of PKS 0903-57}
\label{Sec:Sec7_MW_Study}

 In this section, we have discussed multi-wavelength study of the source PKS 0903-57. From the $\gamma$-ray light curve we have detected different phases of the source. Then we have searched for multi-wavelength data for this source. Here, we have used X-ray, Ultraviolet (UV) and Optical data from Swift-XRT and UVOT (Ultraviolet/Optical telescope) respectively and Radio data collected by ATCA \citep{2020ATel13638....1S}. Only Flare-1B, Flare-IA, Flare-IB and Flare-IC have simultaneous multi-waveband data corresponding to their $\gamma$-ray flaring activity which only spans 4.4 days, 9.2 days, 5.3 days and 10.6 days respectively. Moreover, the number of observations is few in Swift-XRT and Swift-UVOT.

\subsection{Multi-Wavelength Light Curve of PKS 0903-57}

\begin{figure}
	\centering
	\includegraphics[ width=0.94\textwidth,height=0.55\textwidth]{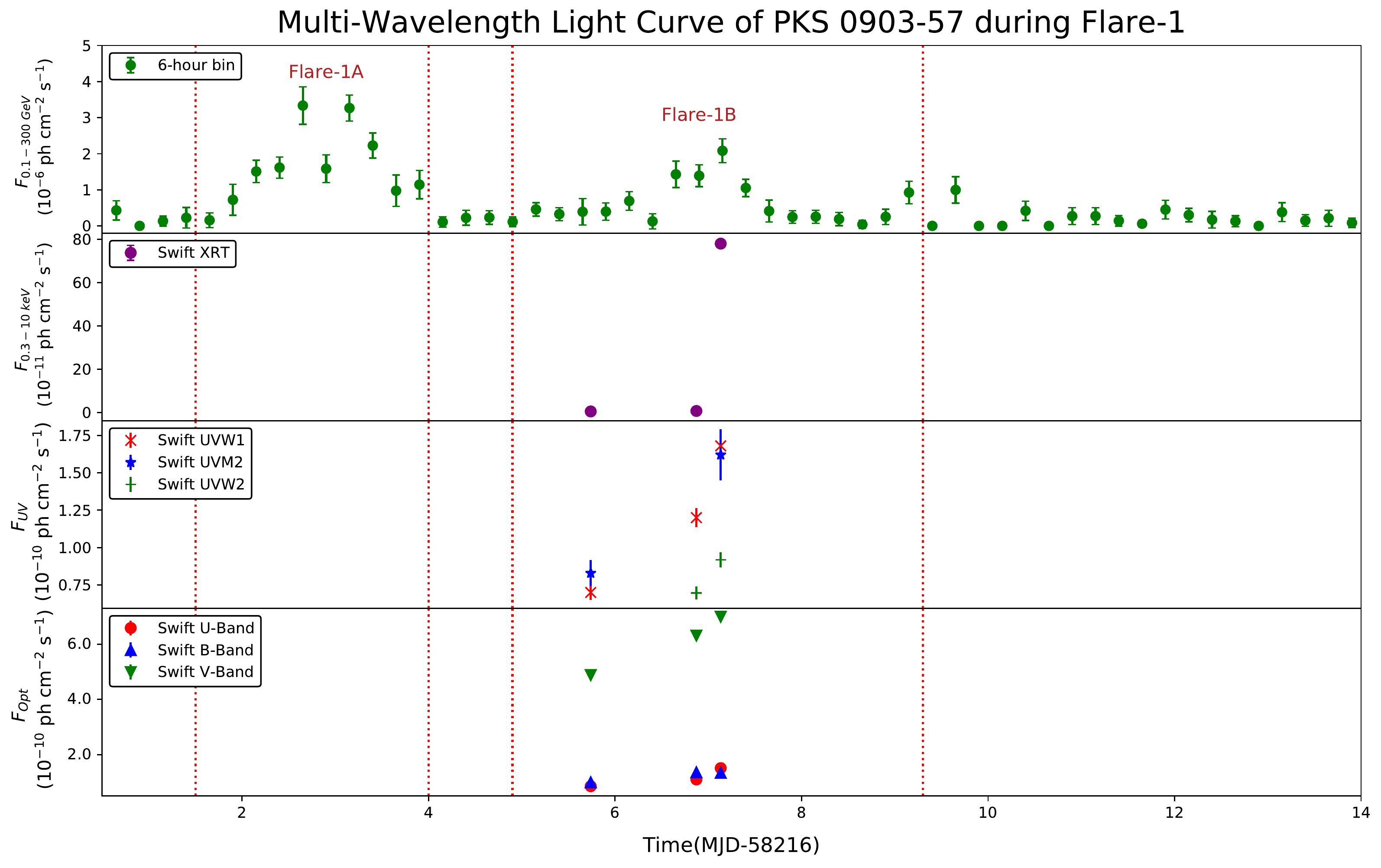}
	\caption{Multi-Wavelength light curve of PKS 0903-57 during Flare-1. The `green solid circle' denotes \textit{Fermi}-LAT data points in 6-hour bin. Others are mentioned in the plots. Ultra-Violet data points are in W1, M2 \& W2 bands and Optical data points are in U, V \& B bands.}
	\label{Fig:MW_LC_Flare-1}
	\end{figure}

 \autoref{Fig:MW_LC_Flare-1} shows the multi-wavelength light curve of the source PKS 0903-57 during Flare-1. Simultaneous multi-wavelength data is only available for Flare-1B, one of the phases of Flare-1, corresponding to MJD 58220.9 to 58225.3 with a period of 4.4 days. In the same plot, We can see that there is no multi-wavelength data corresponding to the $\gamma$-ray light curve of Flare-1A. In the uppermost panel of the plot, 6-hour binned $\gamma$-ray data has been plotted. X-ray, Optical and UV data have been shown in the following panels i.e. in the second, third and fourth panels respectively. We could not get radio data or any data in other wavebands from any other instruments corresponding to Flare-1B. The number of observations in X-ray to Optical is very low to fit the X-ray to Optical light curve and calculate the variability time in X-ray to Optical wavebands. 

\begin{figure}
	\includegraphics[width=0.94\textwidth,height=0.55\textwidth]{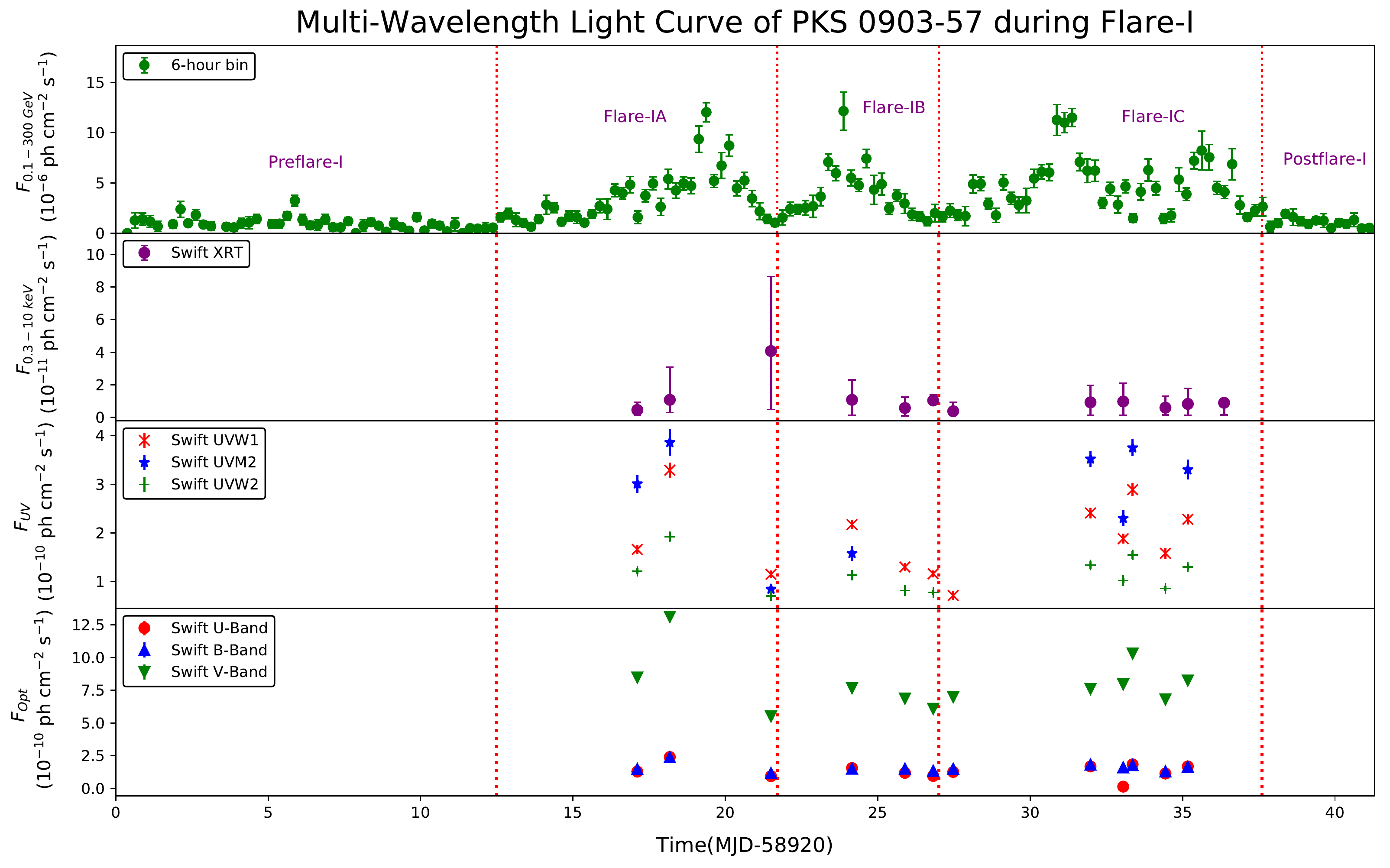}
	\caption{Multi-Wavelength light curve of PKS 0903-57 during Flare-I. Color codes are same as \autoref{Fig:MW_LC_Flare-1}.}
	\label{Fig:MW_LC_Flare-I}
	\end{figure}

In \autoref{Fig:MW_LC_Flare-I}, we have shown multi-wavelength light curve for Flare-I  (MJD 58920.0-58961.3). We have simultaneous multi-wavelength data corresponding to Flare-IA, Flare-IB and Flare-IC i.e. MJD 58932.5-58957.6. Similarly, in the uppermost panel of the plot, 6-hour binned $\gamma$-ray light curve has been shown, followed by X-ray, Optical and UV data in the following panels.

 In Swift-XRT/UVOT we get 15 simultaneous observations corresponding to the flaring states observed in $\gamma$-ray. Out of 15; 3 observations correspond to Flare-1B, 3 observations correspond to Flare-IA, another 3 observations correspond to Flare-IB and the rest of the 6 observations correspond to Flare-IC.

     As we mentioned earlier, the number of observations in X-ray to Optical wavebands is very low; hence it is not possible to do any detail analysis of light curve from X-ray to Optical waveband, only the $\gamma$-ray light curve has been modeled in detail.
     
\subsection{Multi-Wavelength SED Modeling }
We have modeled the multi-wavelength SEDs with a code `GAMERA' \citep{Hahn:2016CO}. It is publicly available on github \footnote{\url{https://github.com/libgamera/GAMERA}} . The code solves time-dependent transport equation. It estimates the propagated electron spectrum N(E,t) for an input injected electron spectrum and further it uses the propagated spectrum to calculate the Synchrotron and Inverse-Compton (IC) emissions. The transport equation used in GAMERA is defined as:
\begin{equation}
\frac{\partial N(E,t)}{\partial t}= Q(E,t)-\frac{\partial}{\partial E}(b(E,t)N(E,t))-\frac{N(E,t)}{\tau\textsubscript{esc}(E,t)}
\end{equation}
where, Q(E,t) is the input electron spectrum and b(E,t) corresponds to the energy loss rate by Synchrotron and IC and can be defined as
\big($\frac{dE}{dt}$\big). In the last term $\tau\textsubscript{esc}$(E,t) denotes the escape time of electrons from the emission region.

Following \cite{2004A&A...413..489M}, a LP photon spectrum can be produced by the radiative losses of a LP electron spectrum. We have considered LP form of injection spectrum. The functional form of the electron spectrum is 

\begin{equation}
Q(E)=L\textsubscript{o}\Bigg(\frac{E}{E_o}\Bigg)^{-\big(\alpha+\beta log\big(\frac{E}{E_o}\big)\big)}
\end{equation}
where L\textsubscript{o} is the normalization constant and E\textsubscript{o} is the scaling factor. This code uses `Klein-Nishina' cross-section to compute Inverse Compton emission \citep{RevModPhys.42.237}.

\subsection{Physical Constraint for Multi-Wavelength SED Modeling}
 
We have used Synchrotron and SSC (Synchrotron Self Compton) emission to model the SEDs. The size of the emission region (R) can be constrained from the causality relation 
\begin{equation}\label{eq:eq_10}
R\leq \frac{c t\textsubscript{var} \delta}{1+z}
\end{equation}
where, t$_{\rm var}$ is the observed variability time, $\delta$ is the Doppler factor of the blob or emission region and $z$ represents the redshift of the source.
We could not find any estimate of Doppler factor ($\delta$) for PKS 0903-57 from earlier studies. The values of Doppler factor for other flaring BL Lacs are found to be in the range of 20 to 40 in most cases. We have used the Doppler factor close to 20 for PKS 0903-57. 
 For Doppler factor 21.5 \& redshift 0.695, the variability time is 1.7$\pm$0.9 hour and the size of the emission region has an upper limit of 2.3$\times$10$^{15}$ cm. But equation \eqref{eq:eq_10} gives only an approximate constraint on the size of the emission region, as there are several other factors that may affect this estimate \citep{protheroe_2002}.

\subsection{Modeling the SEDs}
Varying the fitting parameters in the code `GAMERA' we have modeled multi-wavelength SEDs. In this case, we have considered constant escape of leptons from the emission region with escape time, $\tau_{esc}\sim$R/c, where R is the size of the emission region, used in the fitting and c is the speed of light in vacuum.
\par
\begin{figure}[h]
\centering
\includegraphics[height=9.45cm,width=16.5cm]{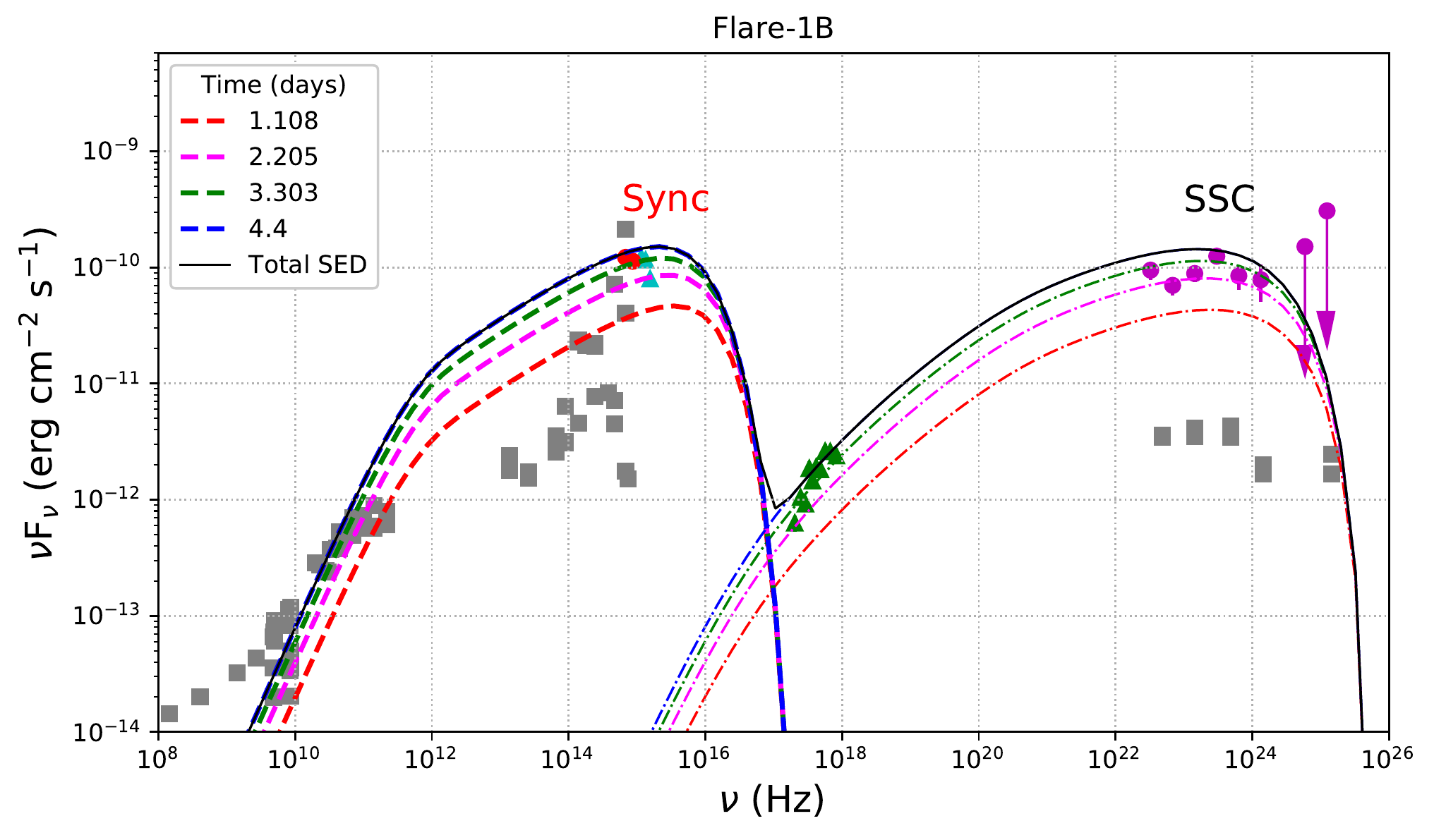}
\caption{Multi-Wavelength SED of Flare-1B. Following are the color codes: `Grey Square'= Archival data points/ Non-simultaneous data points; `Red Solid Circle'= Optical (Swift); `Cyan Triangle'= Ultra-Violet (Swift); `Green Triangle'= X-ray (Swift); `Magenta Solid Circle'= $\gamma$-Ray (\textit{Fermi}-LAT).}
\label{Fig:Flare-1B_GAMERA}
\end{figure}
\begin{figure}[h]
\centering			 
\includegraphics[height=9.45cm,width=16.5cm]{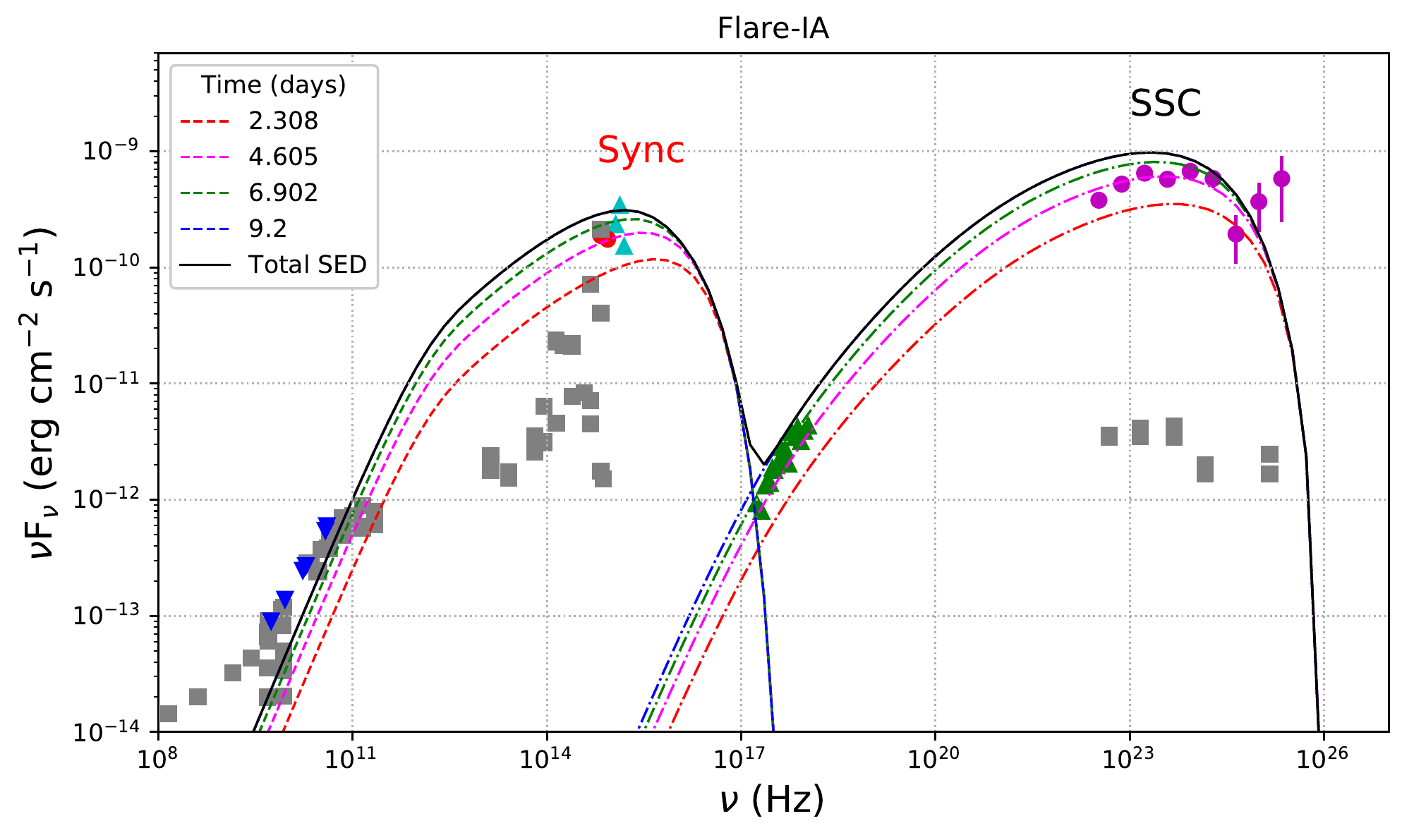}       
\caption{Multi-Wavelength SED of Flare-IA. The color codes are similar to \autoref{Fig:Flare-1B_GAMERA}, radio data points denoted by `Blue inverted-Triangle' (ATCA).}
\label{Fig:Flare-IA_GAMERA}
\end{figure}
\begin{figure}[h]	 
\centering
\includegraphics[height=9.45cm,width=16.5cm]{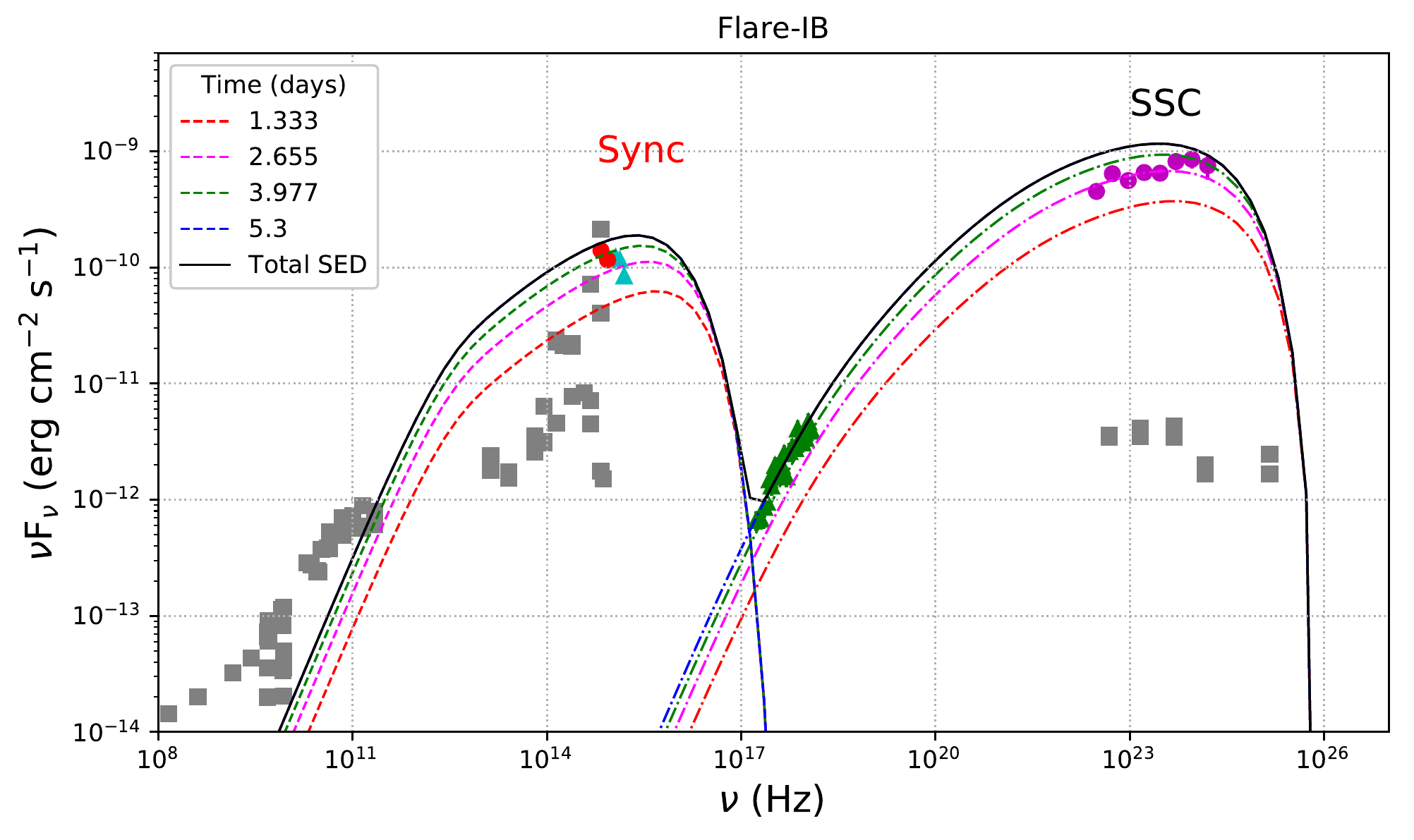}    			\caption{Multi-Wavelength SED of Flare-IB. The color codes are same as \autoref{Fig:Flare-1B_GAMERA}. }
\label{Fig:Flare-IB_GAMERA}
\end{figure}
\begin{figure}[h]
\centering	 
\includegraphics[height=9.45cm,width=16.5cm]{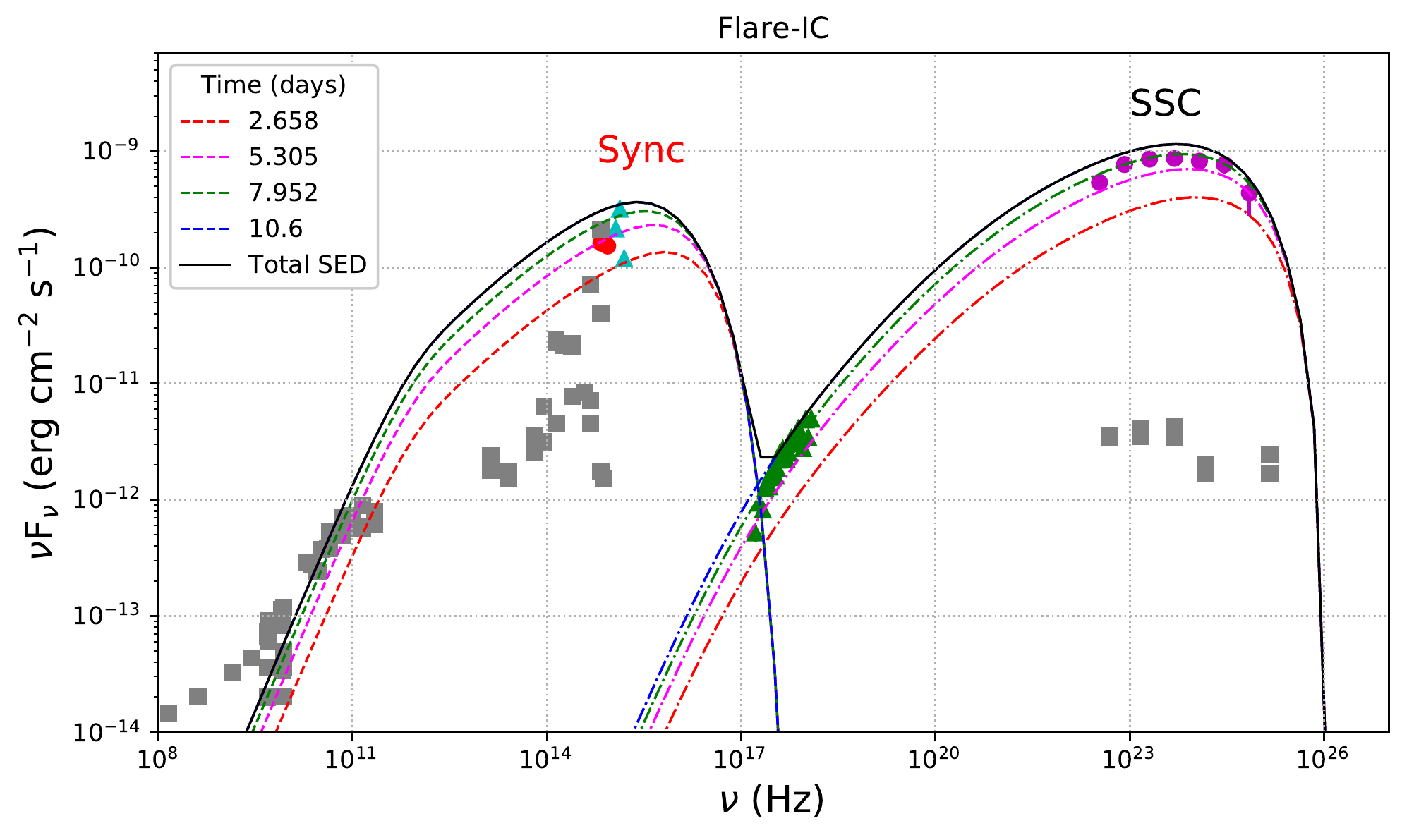}    			\caption{Multi-Wavelength SED of Flare-IC. The color codes are same as \autoref{Fig:Flare-1B_GAMERA}.}
\label{Fig:Flare-IC_GAMERA}  			 
\end{figure}

We have modeled multi-wavelength SEDs of the four phases: Flare-1B, Flare-IA, Flare-IB  and Flare-IC, shown in \autoref{Fig:Flare-1B_GAMERA}, \autoref{Fig:Flare-IA_GAMERA}, \autoref{Fig:Flare-IB_GAMERA}, \& \autoref{Fig:Flare-IC_GAMERA} respectively. For all the phases mentioned above, we have plotted simultaneous data in different wavebands (\textit{Fermi}-LAT $\gamma$-ray: circular magenta points; Swift-XRT: green triangular points; Swift-UV: cyan triangular points; Swift-Optical: red-circular points; ATCA Radio: blue inverted-triangle); also we have shown the non-simultaneous data points in the grey square. We have modeled considering one zone emission region. During the modeling we have adjusted the values of different parameters e.g. minimum and maximum Lorentz factor of the injected electrons ($\gamma$\textsubscript{min} \& $\gamma$\textsubscript{max}), magnetic field (B), size of the emission region (R), spectral index ($\alpha$), curvature index ($\beta$), Doppler factor ($\delta$). All the values of the fitted parameters for the various phases are given in \autoref{tab:GAMERA_Fitting_param}. 
\par

The highest energy photons detected from Flare-1B, Flare-IB \& Flare-IC have energy 5.56 GeV, 6.67 GeV \& 29.33 GeV respectively. 
The three highest energy photons have energy 18.23 GeV, 37 GeV and 81 GeV in Flare-IA.
The optical depth correction due to EBL (Extragalactic Background Light) at redshift 0.695 is negligible for tens of GeV energy \citep{10.1111/j.1365-2966.2012.20841.x} $\gamma$-rays; hence there is no significant attenuation in the SEDs. In Flare-IA (see \autoref{Fig:Flare-IA_GAMERA}), the two highest energy data points show a rising trend in the SED; more observational data points are needed to confirm this trend in future. We have not fitted these two highest energy data points in our model.
\par

 We have also calculated the total jet power using the following equation:
	\begin{equation}
	 P_{tot}=\pi R^2 \Gamma^2 c	(U^\prime_e+U^\prime_B+U^\prime_p)
	 \label{equ:Total_Power}
	\end{equation}
where, $P_{tot}$ is the total jet power; $\Gamma$ is the bulk Lorentz-factor; $U^\prime_e$, $U^\prime_B$ and $U^\prime_p$ are the energy density  of the electrons (and positrons), magnetic field and cold protons respectively in the co-moving jet frame (prime denotes `co-moving jet frame'; unprime denotes `observer frame').
\par
The power carried by the leptons is given by,
\begin{center}
	\begin{equation}
	P_e= \frac{3\Gamma^2c}{4R}  \int_{E_{min}}^{E_{max}} E Q(E) \,dE 
	\end{equation}
\end{center}
where, Q(E) is the injected particle spectrum; integration limits are calculated by multiplying the maximum \& minimum Lorentz factor with the rest-mass energy of electron.
\par

 The power due to magnetic field is calculated by,
	\begin{equation}
	P_B= R^2 \Gamma^2 c \frac{B^2}{8}
	\end{equation}
where, B is the magnetic field, used to model the SED.
\par
The energy density in cold protons $U^\prime_p$ is calculated assuming the number ratio of electron-positron pair to proton is 10:1. We have maintained the charge neutrality condition in the jet. The jet power of protons is computed using the energy density of cold protons.
\par
Subsequently, using \autoref{equ:Total_Power}, we have computed the total jet power of each flaring phases, tabulated in \autoref{Tab:Jet_Power}. We have not found any paper where the mass or the Eddington luminosity of this source is mentioned. The values of jet power reported in \autoref{Tab:Jet_Power} are lower than the typical Eddington luminosities of BL Lacs like Mrk 501, Mrk 421 and AP Librae which are (1.1-4.4)$\times$10$^{47}$ erg/s \citep{Abdo_501}, (2.6-12.0)$\times$10$^{47}$ erg/s \citep{Abdo_421} and 3.75$\times$10$^{46}$ erg/s \citep{Zacharias_2016} respectively.

\section{Summary and Discussion}
\label{Sec:Sec8_Discussion}
\begin{figure}[h]
\centering
\includegraphics[width=0.95\textwidth]{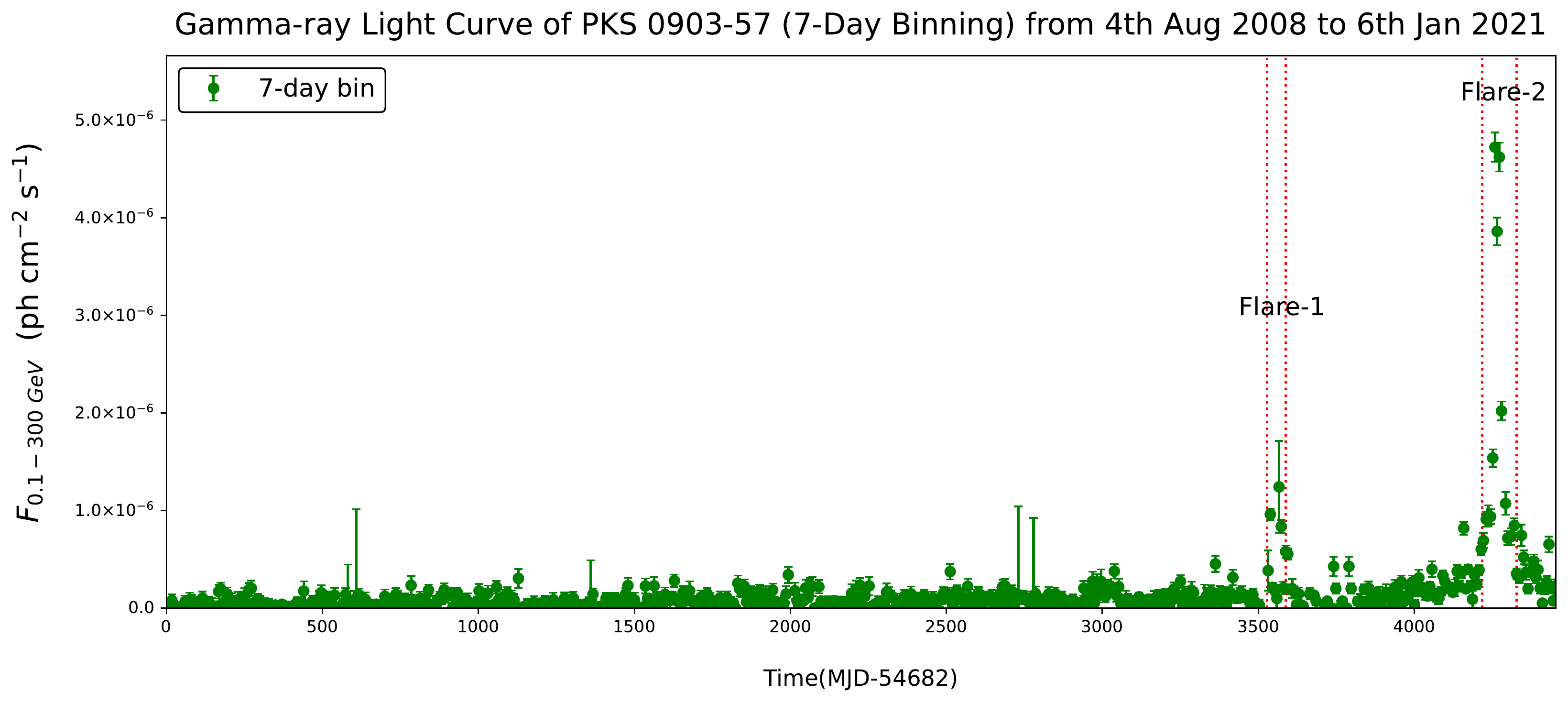}
\caption{12 years (MJD 54682.0-59220.0) $\gamma$-ray light curve of PKS 0903-57 in 7-day binning. Two flaring states Flare-1 \& Flare-2 have been identified and highlighted with a pair of vertical red-dotted lines.}
\label{Fig:7D_Whole}
\end{figure}

\begin{figure}[h]
\centering
\includegraphics[width=0.94\textwidth]{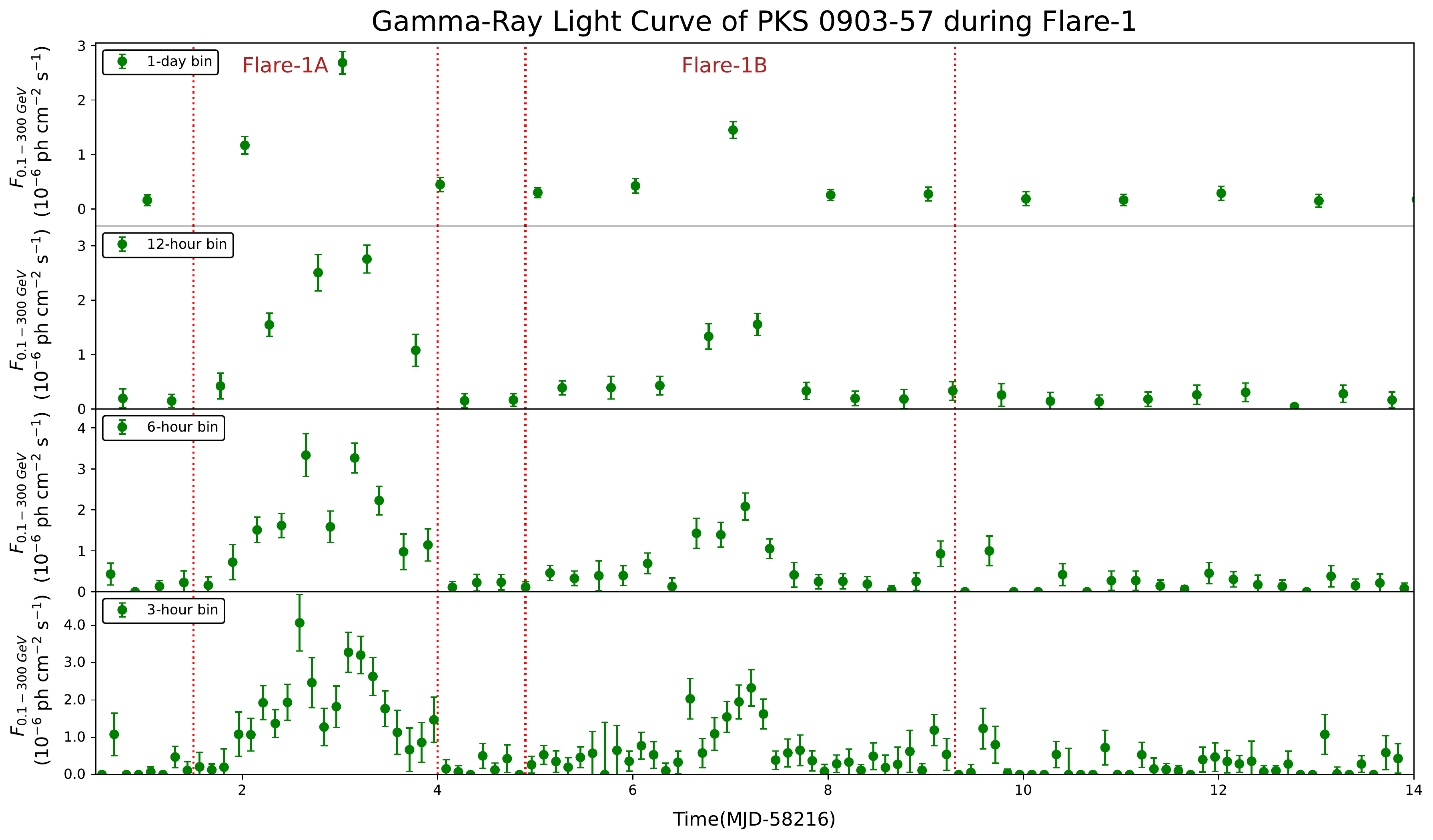}
\caption{$\gamma$-ray light curve of PKS 0903-57 during Flare-1 (MJD 58216.5-58230.0) shown in \autoref{Fig:7D_Whole}. In smaller time binning the different phases of Flare-1 are more prominent. Flare-1 has two phases: Flare-1A \& Flare-1B. }
\label{Fig:Flare-1_Multi_Bin}
\end{figure}

\begin{figure}
\centering
            \begin{minipage}{0.49\textwidth}
 			 \includegraphics[width=0.99\textwidth]{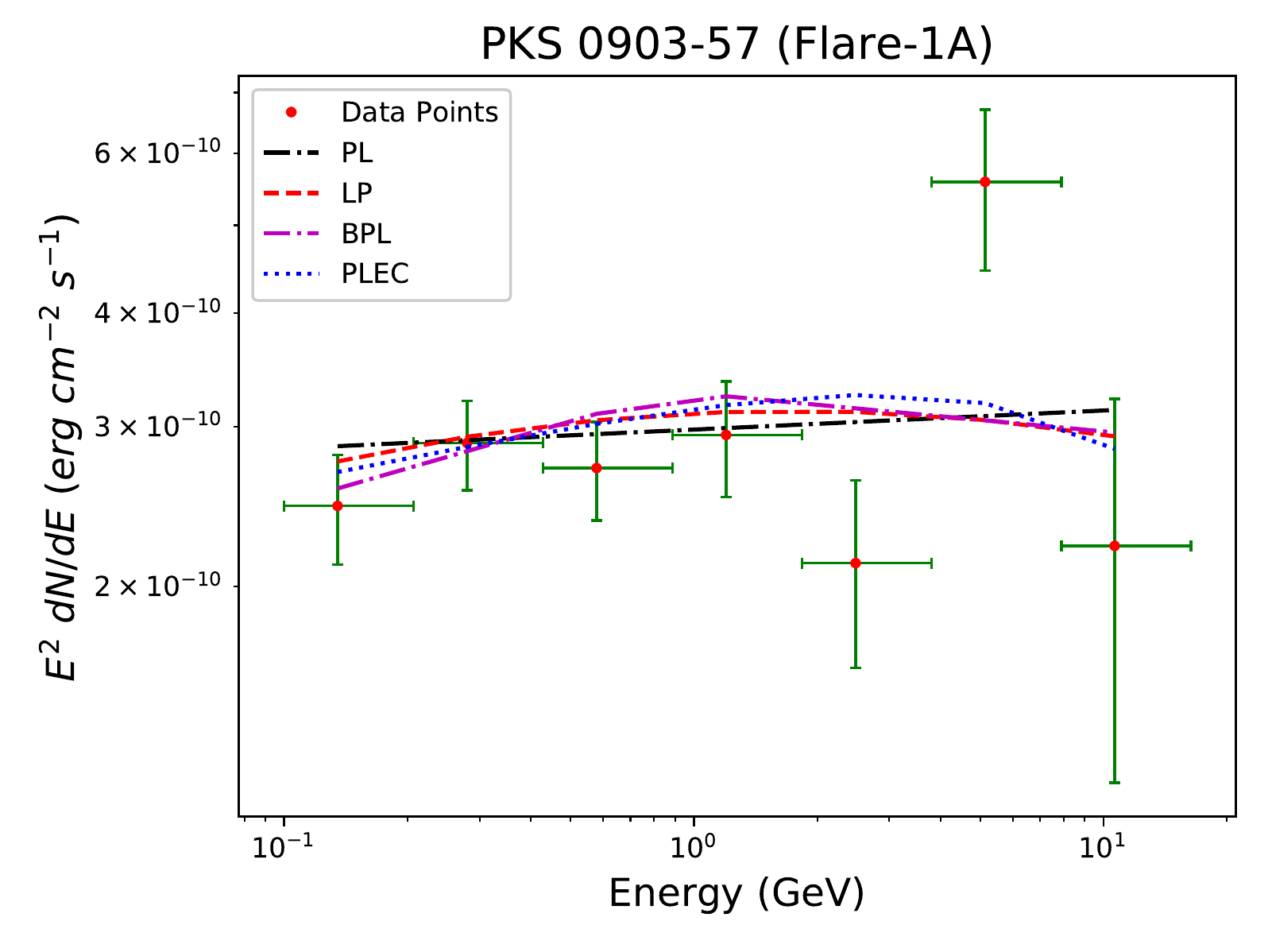}
 			 \end{minipage} \hfill
 			 \begin{minipage}{0.49\textwidth}
 			  \includegraphics[width=0.99\textwidth]{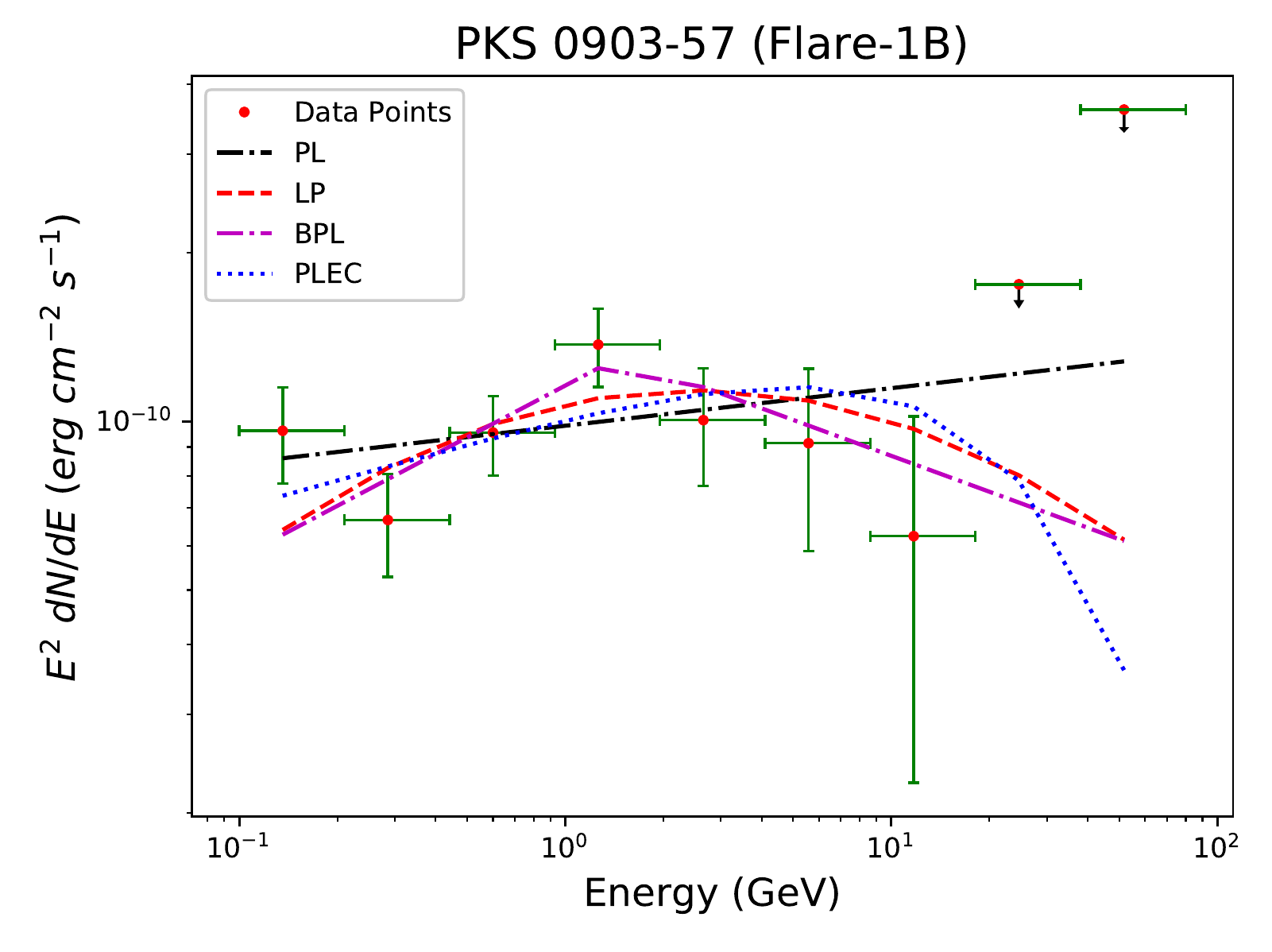}
 			 \end{minipage}
			 \caption{$\gamma$-ray SEDs during Flare-1A \& Flare-1B of Flare-1 as shown in \autoref{Fig:Flare-1_Multi_Bin}. PowerLaw (PL), LogParabola (LP), BrokenPowerLaw (BPL) and PowerLaw with Exponential Cutoff (PLEC) models used to fit the $\gamma$-ray data points. }
             \label{Fig:Flare-1_GammaR_SED}	
\end{figure}

PKS 0903-57 is a BL Lac type blazar, listed in Fermi's regularly monitored source-list and monitored continuously since August 2008. Last year i.e. in 2020, high flaring activity form this source has been detected in different telescopes in different wavebands. Such activity was also reported before e.g. in 2015 and 2018; 7-day binned $\gamma$-ray light curve (\autoref{Fig:7D_Whole}) over the 12 years does not show any significant activity around 2015. Some flaring states are observed in the $\gamma$-ray light curve in 2018 and 2020. We continued our analysis focusing on these activity periods. In \autoref{Fig:7D_Whole}, we have shown 7-day binned $\gamma$-ray light curve over 12 years; from this figure (\autoref{Fig:7D_Whole}), we have denoted two major flaring activities, denoted as Flare-1 \& Flare-2. Further shorter time binning (1-day, 12-hour, 6-hour, 3-hour) reveals sub-structures of these flares. Flare-1 has only one sub-structure whereas Flare-2 has two sub-structures i.e. Flare-I \& Flare-II. In shorter time binned $\gamma$-ray light curve, we have detected different phases (preflare, flare, postflare) of each sub-structure; even several distinctive peaks of each flare region have been detected. Flare-1 has two phases: Flare-1A \& Flare-1B (\autoref{Fig:Flare-1_Multi_Bin}). Flare-1A has two peaks: P$_1$ \& P$_2$ (\autoref{Fig:Flare-1A_LC_Fitting}) and Flare-1B has only one peak: P$_1$ (\autoref{Fig:Flare-1B_LC_Fitting}). The $\gamma$-ray SEDs of Flare-1 have been fitted with PL, LP, BPL and PLEC to check which spectral model gives the best fit to the spectral data (\autoref{Fig:Flare-1_GammaR_SED}). A similar procedure has been followed for the following flares i.e. on Flare-I  \& Flare-II. In most cases, it has been found that the $\gamma$-ray SEDs of the phases can be well described by the LP model. All the calculations done here is based on 3-hour binned $\gamma$-ray light curve. We have calculated the shortest variability time in $\gamma$-ray, which is found to be 1.7$\pm$0.9 hour. We have also studied the rising time ($T_r$) and decay time ($T_d$) of the flaring phases with \autoref{Eqn:LC_Fitting}, to check whether they follow any trend or not. The rising time and decay time have been calculated for each peak, mentioned in \autoref{Tab:Flare-1A_RD_Time} to \autoref{Tab:Flare-II_RD_Time}. The rising and decay timescale found in our study is the order of hour scale. For comparative study, we have considered a quantity $\eta$ \citep{Abdo_2010}. 

\begin{center}
	\begin{equation}
		\eta=\frac{T_d-T_r}{T_d+T_r}
	\end{equation}
\end{center}
 where, -1$< \eta < $1. Depending on the value of the $\eta$, there are three scenarios:
\begin{itemize}

	\item If the rising and decay timescale are nearly equal i.e. T$_r\sim$ T$_d$, symmetric temporal evolution. This can be seen in symmetric flares for which -0.3$< \eta <$0.3.

	\item If rising timescale is greater than the decay timescale i.e. T$_r > $ T$_d$, when $\eta<$-0.3; then injection rate of the electrons is slower than the cooling rate of the electrons into the emission region.
	
	\item If the decay timescale is greater than the rising timescale i.e. T$_d >$ T$_r$, when $\eta>$0.3. This means the electrons take longer time to cool down into the emission region.	
	
\end{itemize}

From our analysis, we found that out of total 15 peaks;  6 peaks have T$_d>$T$_r$, 4 peaks have T$_d<$T$_r$ and 5 peaks have T$_d\sim$T$_r$. It is clear that there is no particular pattern in rising and decay timescale for this source. A flaring part is denoted as a `peak' only when the light curve covered a sufficient number of points; if there are very few points, e.g. 2 or 3, we have not considered them as a `peak'.
\par
Simultaneous multi-wavelength data are available only for four phases: Flare-1B, Flare-IA, Flare-IB  \& Flare-IC in  Swift-XRT, Swift-UVOT and ATCA; though the data in UVOT and Radio are very less. We have modeled these four phases with a time-dependent code `GAMERA'. `GAMERA' solves the transport equation for electrons; it also considers the energy loss by Synchrotron and Synchrotron Self Compton (SSC) process and escapes from the emission region. We have considered a constant escape from the emission region where the escape timescale is R/c $\sim$10$^6$ s. We have modeled with `single-zone' model. The details of the parameters have been mentioned in \autoref{tab:GAMERA_Fitting_param}. We have divided the total flaring duration into four equal time intervals for each of the four phases and we can see the distinct SEDs corresponding to each time interval. The total time duration of Flare-1B, Flare-IA, Flare-IB  and Flare-IC are 4.4 days, 9.2 days, 5.3 days and 10.6 days respectively. 
  \par
 To fit the $\gamma$-ray light curve, we have used \autoref{Eqn:LC_Fitting}; the `Blazar community' uses this function to model the peaks in $\gamma$-ray light curve. The first part of the above-mentioned equation is used to fit the rising part, which gives the rising time. We can estimate the decay timescale by fitting the decaying part of a flare with the second part of the equation. If a flare contains more than one peak, in that case, we have considered a sum of the exponents of the rising \& decay time to fit all the detected peaks in that phase. In this case, peak flux (F$_\circ$) and peak time (t$_\circ$) will be different for different peaks, which is already known from observation. There are several reasons for which the fit may not be good e.g. low TS, fewer data points and large error bars on the data points. In case of rapid flux change, it is difficult to fit all the peaks (even the small peaks) which could be a possible reason behind the poor fitting.
\par
The time binning of the $\gamma$-ray light curve has not been chosen arbitrarily. It is done based on the quality of the data i.e. TS value of each data point. For a very bright $\gamma$-ray source and very high flux, the data may be of very good quality and we can bin the light curve upto minute timescale \citep{Shukla_2018}. In our analysis, we have scanned the 3-hour binned $\gamma$-ray light curve for which TS$\geq$25 i.e. the data points have 5$\sigma$ significance. The $\gamma$-ray flux error increases as the bin size decreases (if we compare 3-hour \& 6-hour time binning, we can easily notice this). Moreover, shorter time binning than 3-hour would be difficult for our analysis. Also, to define a `peak', a fitting curve must cover a sufficient number of points, which is possible if we choose 3-hour time bin instead of 6-hour time bin.
The optical depth correction due to EBL is not important in our case as the energy of the observed highest energy photons is only a few GeV.

\section{Conclusion}
\label{Sec:Sec9_Conclusion}
We have analysed 12 years (From 4th August 2008 to 6th Jan 2021) $\gamma$-ray light curve of PKS 0903-57, from which we have detected two flaring activities in 2018 \& 2020. The $\gamma$-ray flux was the highest in 2020. We have identified two flares: Flare 1 and Flare 2. Flare-1 has one sub-structure, which has two phases: Flare-1A \& Flare-1B. Flare 2 has two sub-structures: Flare I and Flare II, which have several phases. Flare-I has five phases: Preflare-I, Flare-IA, Flare-IB, Flare-IC \& Postflare-I. Flare-II has three phases: Preflare-II, Flare-II \& Postflare-II. We have fitted Flare-1A, Flare-1B, Flare-IA, Flare-IB, Flare-IC \& Flare-II with \autoref{Eqn:LC_Fitting} and calculated the rising and decay time of the peaks of the flaring phases. We have computed the $\gamma$-ray variability time of this source, which is found to be 1.7$\pm$0.9 hour. The different phases of the $\gamma$-ray SEDs have been fitted with PL, LP, BPL \& PLEC to find the best-fitted spectral model. 
 Flare-1B, Flare-IA, Flare-IB \& Flare-IC have simultaneous multi-wavelength data, for these phases LP is found to be the best-fitted model.
 The multi-wavelength SEDs of these four phases have been modeled with a time-dependent code, `GAMERA'. Due to insufficient multi-wavelength data, further multi-wavelength analysis is not possible for this source. We have assumed the emissions are happening from a single-zone. The total jet power required during the flaring phases is estimated to be a few times $10^{46}$ erg/sec.

\section{Software and third party data repository citations} \label{sec:cite}
The \textit{Fermi}-LAT $\gamma$-ray data analysis was done with `Fermi Science Tools software' \citep{Fermi_Sc_Tools}. Swift X-ray, Ultraviolet \& Optical data have been analysed with `Heasoft'.

\acknowledgments
 We thank the referees for insightful comments which improved our work significantly. S.K.M. thanks S. Das, A. Agarwal, T. Ghosh, S. Kabiraj, Hemanth M., A. Dutta, N. N. Patra for useful discussions. R.P. acknowledges the support by the Polish Funding Agency National Science Centre, project 2017/26/A/ST9/00756(MAESTRO 9), and MNiSW grant DIR/WK/2018/12.

%

\facilities{\textit{Fermi}-LAT, Swift(XRT/UVOT), ATCA, DAMPE, AGILE.}


\software{ Fermi Science Tools or Fermitools (\url{https://fermi.gsfc.nasa.gov/ssc/data/analysis/scitools/}) \\  
 Heasoft (\url{https://heasarc.gsfc.nasa.gov/docs/software/lheasoft/})\\
 GAMERA (\url{http://libgamera.github.io/GAMERA/docs/main_page.html})
          }


\vspace{2cm}
\bibliography{PKS0903}{}
\bibliographystyle{aasjournal}



\appendix

\newpage

%


\pagebreak

\pagebreak



\pagebreak

%
%

\pagebreak


\pagebreak

\begin{table}[h]
	\caption{Table for SWIFT XRT/UVOT observations, used in this paper}
	\centering
		\begin{tabular}{|c c c c c c c c c c c  | }
		\hline
		Sr. No. &	& Instrument & & Observation ID & & Starting Time & & XRT Exposure && UVOT Exposure  \\
			&& && && (MJD) && (ks) && (ks) \\
		\hline \hline
		1 && SWIFT XRT/UVOT && 00033856003 && 58221.739 && 1.9 && 1.9 \\
		\hline
		2 && SWIFT XRT/UVOT && 00033856004 && 58222.872 && 1.7 && 1.7 \\
		\hline
		3 && SWIFT XRT/UVOT && 00033856005 && 58223.133 && 1.3 && 1.3 \\
		\hline
		4 && SWIFT XRT/UVOT && 00033856009 && 58937.110 && 1.9 && 1.9 \\
		\hline
		5 && SWIFT XRT/UVOT && 00033856010 && 58938.176 && 1.9 && 1.9 \\
		\hline
		6 && SWIFT XRT/UVOT && 00033856011 && 58941.496 && 1.7 && 1.7 \\
		\hline
		7 && SWIFT XRT/UVOT && 00033856012 && 58944.154 && 2.2 && 2.1 \\
		\hline
		8 && SWIFT XRT/UVOT && 00033856014 && 58945.890 && 1.9 && 1.9 \\
		\hline
		9 && SWIFT XRT/UVOT && 00033856015 && 58946.817 && 1.9 && 1.9 \\
		\hline
		10 && SWIFT XRT/UVOT && 00033856013 && 58947.474 && 0.6 && 0.5 \\
		\hline
		11 && SWIFT XRT/UVOT && 00033856016 && 58951.978 && 1.9 && 1.9 \\
		\hline
		12 && SWIFT XRT/UVOT && 00033856017 && 58953.051 && 1.9 && 1.9 \\
		\hline
		13 && SWIFT XRT/UVOT && 00033856018 && 58954.436 && 0.7 && 0.7 \\
		\hline
		14 && SWIFT XRT/UVOT && 00033856019 && 58955.171 && 2.0 && 1.9 \\
		\hline
		15 && SWIFT XRT/UVOT && 00033856020 && 58956.360 && 1.9 && 1.9 \\
		\hline \hline

		\end{tabular}
		\label{tab:SWIFT Observation IDs}

\end{table}

\begin{center}
	\begin{table*}
	\centering
	\caption{Table of average $\gamma$-ray flux of different activity states of PKS 0903-57}
		\begin{tabular}{||c c c c c ||}
		\hline \hline
		& Activity & Period & Average Gamma-Ray Flux &  \\
		& & (MJD) & (10$^{-6}$ ph cm$^{-2}$  s$^{-1}$) & \\
		\hline \hline
		& Flare-1A & 58217.5-58220.0 & 1.8$\pm$0.5 & \\
		& Flare-1B & 58220.9-58225.3 & 0.6$\pm$0.4 & \\
		\hline
		 &&&&\\
		\hline
		&Preflare-I& 58920.0-58932.5 & 0.9$\pm$0.6 & \\
		& Flare-IA   & 58932.5-58941.7 & 3.6$\pm$1.0 & \\
		& Flare-IB  & 58941.7-58947.0 & 3.9$\pm$1.2 &\\
		& Flare-IC & 58947.0-58957.6 & 4.6$\pm$1.2 & \\
		& Postflare-I & 58957.6-58961.3 & 1.1$\pm$0.7 &\\
		\hline
		&&&&\\
		\hline
		&Preflare-II & 58961.3-58962.0 & 1.2$\pm$0.6 & \\
		& Flare-II & 58962.0-58965.0 & 4.6$\pm$1.0 & \\
		& Postflare-II & 58965.0-58976.0 & 1.1$\pm$0.7 &\\
		\hline \hline
	
		\end{tabular}
	
	\label{Tab:Avg_GammaR_Flux}
	\end{table*}
\end{center}
\begin{center}

\begin{table*}
	\centering
	\caption{Table of Rise \& Decay time for Flare-1A}
		\begin{tabular}{||c c c c c||}
		\hline
		Peak & t$_{\circ}$ & F$_{\circ}$ & $T_{r}$ & $T_{d}$ \\
		 & (MJD) & (10$^{-6}$ ph cm$^{-2}$  s$^{-1}$) & (hr) & (hr)\\
	   \hline\hline
	   $P_{1}$ &  58218.59 & 4.06$\pm$0.75 & 1.8$\pm$0.4 & 2.1$\pm$ 0.5 \\
	   \hline
	   $P_{2}$ &  58219.08 & 3.27$\pm$0.53 &  2.0$\pm$0.6 & 8.0$\pm$ 0.7\\
	   \hline
       \end{tabular}
       \label{Tab:Flare-1A_RD_Time}
\end{table*}       	
       

\begin{table*}	
	\centering
 \caption{Table of Rise \& Decay Time for Flare-1B}
		\begin{tabular}{||c c c c c||}
		\hline
		Peak & t$_{\circ}$ & F$_{\circ}$ & $T_{r}$ & $T_{d}$ \\
		 & (MJD) & (10$^{-6}$ ph cm$^{-2}$  s$^{-1}$) & (hr) & (hr)\\
	   \hline\hline
	   $P_{1}$ &  58223.28 & 2.32$\pm$0.48 & 7.0$\pm$0.9 &  1.8$\pm$ 0.7 \\
	   \hline
	   \end{tabular}
	   \label{Tab:Flare-1B_RD_Time}
\end{table*}	   
	   

\begin{table*}	
	\centering
		\caption{Table of Rise \& Decay Time for Flare-IA}
		\begin{tabular}{||c c c c c||}
		\hline
		Peak & t$_{\circ}$ & F$_{\circ}$ & $T_{r}$ & $T_{d}$ \\
		 & (MJD) & (10$^{-6}$ ph cm$^{-2}$  s$^{-1}$) & (hr) & (hr)\\
	   \hline\hline
	   $P_{1}$ & 58936.90 &  5.10$\pm$0.95 & 13.4$\pm$1.8 & 3.0$\pm$0.9\\
	   \hline
	   $P_{2}$ & 58937.38 &  6.54$\pm$1.08 & 0.5$\pm$0.3 &  3.9$\pm$1.5\\
	   \hline
	   $P_{3}$ & 58938.10 &  6.77$\pm$1.55 & 3.8$\pm$2.0 &  8.2$\pm$1.9 \\
	   \hline
	   $P_{4}$ & 58939.50 & 13.59$\pm$1.37 &  5.8$\pm$1.1 & 1.6$\pm$0.4 \\
	   \hline
	   $P_{5}$ & 58940.22 & 9.86$\pm$1.34 & 5.0$\pm$2.0 &  4.3$\pm$1.8\\
	   \hline
	   \end{tabular}
	   \label{Tab:Flare-IA_RD_Time}
\end{table*}	   
%
\begin{table*}
\centering
\caption{Table of Rise \& Decay Time for Flare-IB }

		\begin{tabular}{||c c c c c||}
		\hline
		Peak & t$_{\circ}$ & F$_{\circ}$ & $T_{r}$ & $T_{d}$ \\
		 & (MJD) & (10$^{-6}$ ph cm$^{-2}$  s$^{-1}$) & (hr) & (hr)\\
	   \hline\hline
	   $P_{1}$ & 58943.76 & 14.13$\pm$2.46 & 1.4$\pm$0.3 & 4.4$\pm$0.6 \\
	   \hline
	   $P_{2}$ & 58944.50 & 7.64$\pm$1.16 & 2.8$\pm$0.8 & 5.4$\pm$1.4 \\
	   \hline
	   \end{tabular}
	   \label{Tab:Flare-IB_RD_Time}
\end{table*}	   
%

\begin{table*}
\centering
\caption{Table of Rise \& Decay Time for Flare-IC}		
	
		\begin{tabular}{||c c c c c||}
		\hline
		Peak & t$_{\circ}$ & F$_{\circ}$ & $T_{r}$ & $T_{d}$ \\
		 & (MJD) & (10$^{-6}$ ph cm$^{-2}$  s$^{-1}$ ) & (hr) & (hr)\\
	   \hline\hline
	   $P_{1}$ & 58948.19 & 5.78$\pm$1.22 & 7.7$\pm$1.5 & 9.2$\pm$1.8\\
	   \hline
	   $P_{2}$ & 58951.31 & 13.44$\pm$1.37 & 15.2$\pm$1.3 & 11.8$\pm$1.2 \\
	   \hline
	   $P_{3}$ & 58953.10 & 7.39$\pm$1.25 & 3.0$\pm$1.0 & 1.3$\pm$0.4\\
	   \hline
	   $P_{4}$ & 58953.75 & 6.84$\pm$1.65 & 2.9$\pm$0.8 & 6.7$\pm$1.1\\
   \hline
	   \end{tabular}
	   \label{Tab:Flare-IC_RD_Time}
\end{table*}	   
	   
	   	   	   	    \vspace*{-1cm}

\begin{table*}
\centering	
	\caption{Table of Rise \& Decay Time for Flare-II }
		\begin{tabular}{||c c c c c||}
		\hline
		Peak & t$_{\circ}$ & F$_{\circ}$ & $T_{r}$ & $T_{d}$ \\
		 & (MJD) & (10$^{-6}$ ph cm$^{-2}$  s$^{-1}$) & (hr) & (hr)\\
	   \hline\hline
	   $P_{1}$ & 58962.94 & 7.78$\pm$1.01 & 3.2$\pm$0.5 & 3.0$\pm$1.4\\
	   \hline
	   \end{tabular}
	   \label{Tab:Flare-II_RD_Time}
\end{table*}

	  \end{center}
\pagebreak


\begin{center}

\begin{table*}
\centering
\caption{Table for $\gamma$-ray flux doubling/halving time ($T_{d/h}$) for each flare}
\begin{tabular}{||c c c c c c c ||}
						 \hline
							 T\textsubscript{start}(t\textsubscript{1}) & T\textsubscript{stop}(t\textsubscript{2}) & Flux\textsubscript{start}[F(t\textsubscript{1})] & Flux\textsubscript{stop}[F(t\textsubscript{2})] & $T_{d/h}$ & $\triangle$t\textsubscript{d/h} & Rise/Decay \\
							(MJD) & (MJD) & (10\textsuperscript{-6} ph cm\textsuperscript{-2} s\textsuperscript{-1}) & (10\textsuperscript{-6} ph cm\textsuperscript{-2} s\textsuperscript{-1}) & (hr) & (hr) &\\
							 \hline \hline
							 &&&&&&\\
					   & & &	 Flare-1 &&& \\
					     \hline
					     58218.463 & 58218.588 & 1.93$\pm$0.47 & 4.06$\pm$0.75 & 2.8$\pm$1.2 & 1.6$\pm$0.7 &  R\\
						 \hline\hline
						 &&&&&&\\
					 & & &	 Flare-I  &&& \\
						 \hline 
					    58922.438 & 58922.563 & 1.26$\pm$0.53 & 2.94$\pm$0.95 &   2.5$\pm$1.6 & 1.5$\pm$0.9 &  R\\    	                \hline
					    58934.313 & 58934.438 & 1.51$\pm$0.68 &  3.19$\pm$0.67 & 2.8$\pm$1.9 & 1.6$\pm$1.1 & R\\    
					     \hline
					    58935.563 & 58935.688 & 2.61$\pm$0.68 & 1.02$\pm$0.50 &  -2.2$\pm$1.3 & -1.3$\pm$0.8 & D\\   
					     \hline
					    58935.688 &  58935.813 & 1.02$\pm$0.50 & 3.45$\pm$1.23 & 1.7$\pm$ 0.9 & 1.0$\pm$0.5 & R\\   
					      \hline
					    58939.563 & 58939.688 & 7.50$\pm$1.07 &  2.97$\pm$0.66 &  -2.2$\pm$ 0.6 & -1.3$\pm$0.4 & D\\  
					      \hline
					    58940.188 & 58940.313 & 9.86$\pm$1.34 & 4.45$\pm$0.98 & -2.6$\pm$0.8 & -1.5$\pm$0.5 & D\\     
					      \hline
					    58940.688 & 58940.813 & 5.77$\pm$1.18 & 2.60$\pm$0.82 & -2.6$\pm$1.2 & -1.5$\pm$0.7 & D\\ 
					      \hline
					    58940.813 & 58940.938 & 2.60$\pm$0.82 & 5.98$\pm$2.01 & 2.5$\pm$1.4 & 1.5$\pm$0.8 & R\\
					     \hline
					     58943.313 & 58943.438 & 4.60$\pm$1.03 & 9.69$\pm$1.28 & 2.8$\pm$ 1.0 & 1.6$\pm$0.6 & R\\  
					     \hline
					    58943.438 & 58943.563 & 9.69$\pm$1.28 & 4.60$\pm$0.98 & -2.8$\pm$0.9 & -1.6$\pm$0.6 & D\\   
					     \hline
					     58945.188 & 58945.313 & 5.91$\pm$1.91 & 2.23$\pm$0.82 & -2.1$\pm$1.1 & -1.3$\pm$0.6 & D\\   
					     \hline
 					    58949.063 &  58949.188 & 7.78$\pm$1.56 & 3.74$\pm$0.79 & -2.8$\pm$1.1 & -1.7$\pm$0.7 & D\\  
					     \hline
					    58953.063 & 58953.188 & 7.39$\pm$1.25 & 2.94$\pm$0.69 & -2.2$\pm$0.7 & -1.3$\pm$0.4 & D\\ 
					     \hline
					     58953.563 & 58953.688 & 2.37$\pm$0.83 &  7.12$\pm$1.64 & 1.9$\pm$0.7 & 1.1$\pm$0.4 & R\\ 
					     \hline
					     &&&&&&\\
					   & & &	 Flare-II  &&&\\ 
					     \hline
					     58962.688 & 58962.813 & 1.82$\pm$0.92 & 3.82$\pm$1.29 & 2.8$\pm$2.3 & 1.7$\pm$1.4 & R \\ 
					     \hline
					    58962.813 & 58962.938 & 3.82$\pm$1.29 & 7.78$\pm$1.01 & 2.9$\pm$1.5 & 1.7$\pm$0.9 & R\\  
					    \hline
					    58963.688 & 58963.813 & 3.32$\pm$1.11 & 8.01$\pm$2.18 & 2.4$\pm$1.2 & 1.4$\pm$0.7 & R\\ 
					    \hline
					     58963.938 & 58964.063 & 6.35$\pm$0.93 & 2.99$\pm$0.62 & -2.8$\pm$0.9 & -1.6$\pm$0.6 & D\\ 
					    \hline
					    58964.313 & 58964.438 & 3.53$\pm$0.71 & 1.56$\pm$0.57 & -2.5$\pm$1.3 & -1.5$\pm$0.8 & D\\
						\hline \hline

\end{tabular}
\tablecomments { $\triangle$t\textsubscript{d/h} is redshift corrected doubling/halving time. `R' denotes `rising part' \& `D' denotes `decay part'.}
\label{Tab:Flux_DH_Time}
\end{table*}

\end{center}



\begin{table*}
\centering
\caption{ Results of \textit{Fermi}-LAT SEDs of Flare-1, fitted with different spectral model e.g. PL, LP, BPL and PLEC}


\begin{tabular}{||c c c c c c c c c c || }
						 \hline
					&& PowerLaw (PL)&&& &&&&\\
						 \hline
							 Activity & F$_{0.1 - 300 \ GeV}$  &  Index ($\Gamma$) &   & & TS & & -log(Likelihood) &  &\\ 
							 &($10^{-6} \ ph \ cm^{-2} \ s^{-1}$)&&&&&&&&\\ 
						 \hline\hline
					     Flare-1A & 1.76$\pm$0.12 & 1.98$\pm$0.05 &-&-& 1288.42 &-& 14395.21 &-& \\
						 \hline
							 Flare-1B & 0.56$\pm$0.06 & 1.93$\pm$0.07 &-&-& 661.70 &-& 34126.0 & -&  \\												 \hline\hline
						 			 
\end{tabular}

\vspace{1.5cm}


\begin{tabular}{||c c c c c  c c ||}
						 \hline
					&&& LogParabola (LP)&&& \\
						 \hline
							 Activity &  F$_{0.1 - 300 \ GeV}$  & $\alpha$ & $\beta$ &  TS &   -log(Likelihood)  & $\triangle$ log(Likelihood) \\
							 & ($10^{-6} \ ph \ cm^{-2} \ s^{-1}$) & & & & &   \\
						 \hline\hline
	     					 Flare-1A & 1.73$\pm$0.14 & 1.98$\pm$0.05 & 0.04$\pm$ 0.02 & 1285.61 & 14394.96 & -0.25\\
						 \hline
							 Flare-1B & 0.48$\pm$0.08 & 1.90$\pm$0.08 & 0.07$\pm$0.05 & 652.83 & 34124.99 & -1.02 \\
						  \hline\hline
						 			 
\end{tabular}

\vspace{0.5cm}
	

\begin{tabular}{||c c c c c c c  c ||}
						  \hline
					&&& BrokenPowerLaw (BPL)&&&& \\
						 \hline
							 Activity &  F$_{0.1 - 300 \ GeV}$   & $\Gamma_{1}$ & $\Gamma_{2}$ & E$_{b}$  & TS & -log(Likelihood) & $\triangle$ log(Likelihood) \\
							 & ($10^{-6} \ ph \ cm^{-2} \ s^{-1}$) & & &(GeV)  & & &\\
						 \hline\hline
						\textit{Flare-1A} & \textit{1.65$\pm$0.11} & \textit{1.87$\pm$0.05} &  \textit{2.04$\pm$0.07} & \textit{0.90$\pm$0.02} & \textit{1292.18} & \textit{14386.14} & \textit{-9.07} \\
						 \hline
							\textit{ Flare-1B} & \textit{0.48$\pm$0.16} & \textit{1.69$\pm$0.34} & \textit{2.21$\pm$0.23} & \textit{1.47$\pm$0.76} & \textit{654.10} & \textit{34123.84} & \textit{-2.17}\\
 						 \hline\hline
						 			 
\end{tabular}

 \vspace{.5cm}
	
\begin{tabular}{||c c c c  c c  c||}
						\hline
						&&& PL Exp Cutoff (PLEC)
&&& \\
						 \hline
							 Activity &  F$_{0.1 - 300 \ GeV}$  & $\Gamma_{PLEC}$ &  E$_{c}$ &TS  & -log(Likelihood) & $\triangle$ log(Likelihood) \\
							 & ($10^{-6} \ ph \ cm^{-2} \ s^{-1}$) & & (GeV) & & &\\
						 \hline\hline
	                         Flare-1A & 1.71$\pm$0.13 & 1.91$\pm$0.06 &  29.88$\pm$12.97 & 1284.82 & 14394.00 & -1.21\\
						 \hline
							 Flare-1B & 0.52$\pm$0.06 & 1.83$\pm$0.07 &  29.99$\pm$0.76 & 656.28  & 34125.73 & -0.28\\
						 \hline\hline
						 			 
\end{tabular}
\label{Tab:Flare-1_GammaR_SED_Param}

\tablecomments{ Here we have mentioned the fitted flux and spectral indices. We have also mentioned the goodness of unbinned fits by -log(Likelihood) value and evaluate $\triangle$log(Likelihood) value, where $\triangle$log(Likelihood)=(-log(Likelihood)$_{LP/BPL/PLEC}$)-(-log(Likelihood)$_{PL}$). \\
The best-fitted models are hightlighted in `\textit{italic}' font.}
\end{table*}



	\begin{table*}
	\centering
	\caption{Results of \textit{Fermi}-LAT SEDs of Flare-I, fitted with different spectral model e.g. PL, LP, BPL and PLEC}
\centering
\begin{tabular}{||c c c c c c c c c c ||}
						 \hline
						 && PowerLaw (PL) &&&&&&& \\
						 \hline
							 Activity & F$_{0.1 - 300 \ GeV}$  &  Index ($\Gamma$) &   & & TS & & -log(Likelihood) &  &\\ 
							 &($10^{-6} \ ph \ cm^{-2} \ s^{-1}$)&&&&&&&&\\ 
						 \hline\hline
	Preflare-I& 0.56 $\pm$ 0.05 & 2.08$\pm$0.06 & - & - & 690.73 & - & 47426.03 &  - &\\ 
						 \hline
						     Flare-IA   & 3.02$\pm$0.11 & 1.91$\pm$0.02 &-&-& 4980.58 &-& 30625.17 &-& \\
						 \hline
							 Flare-IB  & 3.86$\pm$0.16 & 1.94$\pm$ 0.03 &-&-& 3196.55 &-& 19383.86 & -& \\
						 \hline
							 Flare-IC & 4.34$\pm$0.12 & 1.90$\pm$0.02 &-&-& 8505.55 &-& 43270.09 &-& \\ 
						 \hline
						 	Postflare-I & 0.77$\pm$0.11 & 2.08$\pm$0.10 &-&-& 258.35 & - & 11551.58 & - &\\ 
						 \hline\hline
						 			 
\end{tabular}
\vspace{.2cm}


\begin{tabular}{||c c c c c c c||}
						 \hline
						 && &   LogParabola (LP) &&& \\
						 \hline
							Activity &  F$_{0.1 - 300 \ GeV}$  & $\alpha$ & $\beta$ &  TS  & -log(Likelihood)  & $\triangle$ log(Likelihood) \\
							 & ($10^{-6} \ ph \ cm^{-2} \ s^{-1}$) & & &  & & \\
						 \hline\hline
	                        Preflare-I& 0.50 $\pm$ 0.01 & 2.11$\pm$0.00 & 0.11$\pm$0.00 & 712.46  & 47412.73 & -13.30  \\
						  \hline
							\textit{Flare-IA}  &\textit{2.78$\pm$0.12} &\textit{1.91$\pm$0.03} &\textit{0.07$\pm$ 0.02} &\textit{4956.73} &\textit{30613.75} &\textit{-11.42} \\
						 \hline
							\textit{Flare-IB}  &\textit{3.51$\pm$0.17} &\textit{1.93$\pm$0.04} &\textit{0.06$\pm$0.02} &\textit{3055.13 }&\textit{19372.94} & \textit{-10.92} \\
						 \hline
							\textit{Flare-IC } &\textit{4.12$\pm$0.12} &\textit{1.93$\pm$0.02} &\textit{0.07$\pm$0.01} &\textit{8478.70} &\textit{43257.17} &\textit{-12.92} \\
						  \hline
				        	  Postflare-I & 0.72$\pm$0.12 & 2.14$\pm$0.12 & 0.08$\pm$0.07 & 259.05  & 11550.82 & -0.76 \\
						 \hline\hline
						 			 
    \end{tabular}

\vspace{.2cm}
	
\begin{tabular}{||c c c c c c c c ||}
						 \hline
						&&& BrokenPowerLaw (BPL) &&&& \\
						 \hline
							 Activity &  F$_{0.1 - 300 \ GeV}$   & $\Gamma_{1}$ & $\Gamma_{2}$ & E$_{b}$  & TS & -log(Likelihood) & $\triangle$ log(Likelihood) \\
							 & ($10^{-6} \ ph \ cm^{-2} \ s^{-1}$) & & &(GeV)  & & &\\ 
						 \hline\hline
	Preflare-I& 0.50$\pm$0.14 & 1.85$\pm$0.32 & 2.41$\pm$0.22 & 1.16$\pm$0.49 &  686.42  & 47423.08 &  -2.95 \\
						  \hline
							 Flare-IA   & 2.80$\pm$0.15 & 1.72$\pm$0.06 &  2.12$\pm$0.06 & 1.03$\pm$0.18 & 4954.03 & 30614.38 &  -10.79\\
						 \hline
							 Flare-IB  & 3.78$\pm$0.17 & 1.87$\pm$0.04 & 2.16$\pm$0.11 & 2.00$\pm$0.00 & 3201.32 & 19381.47 & -2.39 \\
						 \hline
							 Flare-IC & 4.14$\pm$0.12 & 1.74$\pm$0.04 & 2.10$\pm$0.05 & 0.98$\pm$0.09 & 8481.95  &  43257.18 & -12.91\\
						 \hline
						     Postflare-I & 0.74$\pm$0.25 & 1.94$\pm$0.43 & 2.29$\pm$0.30 & 1.00$\pm$0.68 & 258.46 & 11551.07 & -0.31\\
						 \hline\hline
						 			 
\end{tabular}
	
	
\vspace{0.2cm}
	

\begin{tabular}{||c c c c c   c c||}
						 \hline
						&&& PL Exp Cutoff (PLEC) &&& \\
						 \hline
							  Activity &  F$_{0.1 - 300 \ GeV}$  & $\Gamma_{PLEC}$ &  E$_{c}$ &TS  & -log(Likelihood) & $\triangle$ log(Likelihood) \\
							 & ($10^{-6} \ ph \ cm^{-2} \ s^{-1}$) & & (GeV) & & &\\
						 \hline\hline
	                        Preflare-I& 0.52$\pm$0.03 &  1.89$\pm$0.03 &  9.64$\pm$1.78 & 712.33 & 47416.66 & -9.37 \\
						 \hline
						 Flare-IA   & 2.90$\pm$0.11 & 1.82$\pm$0.03 & 30.00$\pm$0.03 & 4961.45 & 30620.54 & -4.63\\
						 \hline
							 Flare-IB  & 3.75$\pm$0.16 & 1.84$\pm$0.05 & 17.98$\pm$7.06 & 3205.43 & 19379.42 &  -4.44 \\
						 \hline
							 Flare-IC & 4.22$\pm$0.12 & 1.82$\pm$0.03  & 23.38$\pm$6.20 & 8494.66  & 43261.78 &  -8.31\\ 
						 \hline
						 	 Postflare-I & 0.76$\pm$0.11 & 2.02$\pm$0.10  & 30.00$\pm$1.83 & 258.11  & 11551.48 & -0.10 \\
						 \hline\hline
						 			 
\end{tabular}
 \label{Tab:Flare-I_GammaR_SED_Param}	
\end{table*}






\begin{table*}
	
\caption{Results of \textit{Fermi}-LAT SEDs of Flare-II, fitted with different spectral model e.g. PL, LP, BPL, PLEC}
\centering
\vspace{0.1cm}

\begin{tabular}{||c c c c c c c c c c ||}
						\hline
						&& PowerLaw (PL) &&&&&&& \\
						 \hline
							Activity & F$_{0.1 - 300 \ GeV}$  &  Index ($\Gamma$) &   & & TS & & -log(Likelihood) &  &\\ 
							 &($10^{-6} \ ph \ cm^{-2} \ s^{-1}$)&&&&&&&&\\ 
						 \hline\hline
	                    Preflare-II &  0.92$\pm$0.15 & 1.82$\pm$0.10 & - & - & 232.32 & - & 10495.42 &  - &\\ 
						\hline
						 	 Flare-II & 3.19$\pm$0.16 & 1.92$\pm$0.03 &-&-& 2288.81 & - & 15193.42 &-&\\ 
						 	 \hline
						 	 Postflare-II & 0.91$\pm$0.06 & 2.15$\pm$0.05 &-&-& 1166.98 & - & 49385.92 &-&\\ 
						 \hline\hline
						 			 
\end{tabular}


	


\begin{tabular}{||c c c c c c  c||}
						 \hline
						 &&& LogParabola (LP) &&& \\
						 \hline
							 	Activity &  F$_{0.1 - 300 \ GeV}$  & $\alpha$ & $\beta$ &  TS  & -log(Likelihood) & $\triangle$ log(Likelihood)  \\
							 & ($10^{-6} \ ph \ cm^{-2} \ s^{-1}$) & & &  &  & \\
						 \hline\hline
	Preflare-II & 0.85 $\pm$ 0.02 & 1.78$\pm$0.01 & 0.02$\pm$0.01 & 244.00 & 10450.08 & -45.34   \\
						 \hline
							Flare-II & 3.01$\pm$0.16 & 1.94$\pm$0.04 & 0.08$\pm$0.02 & 2302.13  & 15186.76 &  -6.66\\
							\hline
							Postflare-II & 0.84$\pm$0.07 & 2.21$\pm$0.06 & 0.09$\pm$0.04 & 1156.62  & 49376.90 & -9.02\\
						 \hline\hline
						 			 
\end{tabular}
		
	
	


\begin{tabular}{||c c c c c c c  c ||}
						 \hline
						 &&& BrokenPowerLaw (BPL) &&&& \\
						 \hline
							 Activity &  F$_{0.1 - 300 \ GeV}$   & $\Gamma_{1}$ & $\Gamma_{2}$ & E$_{b}$  & TS & -log(Likelihood) & $\triangle$ log(Likelihood) \\
							 & ($10^{-6} \ ph \ cm^{-2} \ s^{-1}$) & & &(GeV)  & & &\\ 
						 \hline\hline
	Preflare-II & 0.78$\pm$0.00 & 1.36$\pm$0.00 & 1.95$\pm$0.00 & 0.61$\pm$0.00 &  244.93  & 10449.71 & -45.71\\
						  \hline
							\textit{Flare-II} &\textit{2.99$\pm$0.16} &\textit{1.62$\pm$0.11}  &\textit{2.13$\pm$0.08} &\textit{0.65$\pm$0.20} &\textit{2304.57} &\textit{15185.54} &\textit{-7.88}\\
							\hline
							Postflare-II & 0.86$\pm$0.07 & 2.00$\pm$0.08 & 2.41$\pm$0.13 & 1.07$\pm$0.11 & 1158.88 & 49377.54 & -8.38 \\
						 \hline\hline
						 			 
\end{tabular}


	

\begin{tabular}{||c c c c c  c  c||}
						\hline
						&&& PL Exp Cutoff (PLEC) &&& \\
						 \hline
							 Activity &  F$_{0.1 - 300 \ GeV}$  & $\Gamma_{PLEC}$ &  E$_{c}$ &TS  & -log(Likelihood) & $\triangle$ log(Likelihood) \\
							 & ($10^{-6} \ ph \ cm^{-2} \ s^{-1}$) & & (GeV) & & &\\
						 \hline\hline
	                        Preflare-II & 0.88$\pm$0.15 & 1.72$\pm$0.11  & 30.00$\pm$0.03 & 228.14  & 10497.51 &  2.09\\
						 \hline
							 Flare-II & 3.11$\pm$0.16 & 1.83$\pm$0.05  & 26.53$\pm$12.49 & 2296.64  & 15189.51 &  -3.91\\
							 \hline
							 Postflare-II & 0.87$\pm$0.06 & 2.03$\pm$0.08 & 13.94$\pm$8.04 & 1163.43  & 49377.84 &  -8.08\\
						 \hline\hline
						 			 
\end{tabular}
\label{Tab:Flare-II_GammaR_SED_param}
 \end{table*}

\pagebreak

\begin{table*}
\caption{Results of multi-wavelength SED modeling shown in the \autoref{Fig:Flare-1B_GAMERA} to  \autoref{Fig:Flare-IC_GAMERA}}
\centering
\begin{tabular}{cccc rrrr}   
\hline\hline
 Parameters & Symbol & values & Time duration  \\
\hline
& Flare-1B \\
\hline
 Spectral index of injected electron spectrum (LP) & $\alpha$ & 2.1 \\
  Curvature index of injected electron spectrum & $\beta$ & 0.09 \\

 Magnetic field in emission region & B & 0.25 G  \\
 Size of the emission region & R & 6.6$\times10^{16}$ cm  & 4.4 days \\
 Doppler factor of emission region & $\delta$ & 21.5 \\

 Min. value of Lorentz factor of injected electrons & $\gamma_{min}$ & 1.5$\times10^2$  \\
 Max. value of Lorentz factor of injected electrons & $\gamma_{max}$ & 3.0$\times10^4$  \\

\hline
& Flare-IA   \\
\hline
 Spectral index of injected electron spectrum (LP) & $\alpha$ & 1.7 \\
  Curvature index of injected electron spectrum & $\beta$ & 0.20 \\

 Magnetic field in emission region & B & 0.25 G  \\
 Size of the emission region & R & 5.9$\times10^{16}$ cm  & 9.2 Days \\
 Doppler factor of emission region & $\delta$ & 21.5 \\
 
 Min. value of Lorentz factor of injected electrons & $\gamma_{min}$ & 2.4$\times10^2$  \\
 Max. value of Lorentz factor of injected electrons & $\gamma_{max}$ & 4.5$\times10^4$  \\
\hline
& Flare-IB  \\
\hline
 Spectral index of injected electron spectrum (LP) & $\alpha$ & 1.7 \\
  Curvature index of injected electron spectrum & $\beta$ & 0.20 \\

 Magnetic field in emission region & B & 0.25 G  \\
 Size of the emission region & R & 3.0$\times10^{16}$ cm  & 5.3 Days \\
 Doppler factor of emission region & $\delta$ & 21.5 \\

 Min. value of Lorentz factor of injected electrons & $\gamma_{min}$ & 3.4$\times10^2$  \\
 Max. value of Lorentz factor of injected electrons & $\gamma_{max}$ & 3.9$\times10^4$  \\
\hline
& Flare-IC \\
\hline
 Spectral index of injected electron spectrum (LP) & $\alpha$ & 1.7 \\
  Curvature index of injected electron spectrum & $\beta$ & 0.17 \\

 Magnetic field in emission region & B & 0.19 G  \\
 Size of the emission region & R & 8.0$\times10^{16}$ cm  & 10.6 Days \\
 Doppler factor of emission region & $\delta$ & 21.5 \\

 Min. value of Lorentz factor of injected electrons & $\gamma_{min}$ & 2.0$\times10^2$  \\
 Max. value of Lorentz factor of injected electrons & $\gamma_{max}$ & 5.5$\times10^4$  \\
\hline

\end{tabular}
\label{tab:GAMERA_Fitting_param}
\end{table*}

\begin{table*}
\centering
\caption{Table for total jet power for each flaring event}
	\begin{tabular}{||ccc c c c c||}
		\hline
		&&Activity & & Total Jet Power&&\\
		&&& & (erg/s)&& \\
		\hline
		&&Flare-1B & & 1.3$\times$10$^{46}$ &&\\
		\hline
		&&Flare-IA & & 2.3$\times$10$^{46}$&&\\
		\hline 
		&&Flare-IB & & 3.0$\times$10$^{46}$&&\\
		\hline
		&&Flare-IC & & 1.2$\times$10$^{46}$&&\\
		\hline
	
	\end{tabular}
	\label{Tab:Jet_Power}

\end{table*}

\end{document}